\documentclass[twocolumn]{aastex7}

\usepackage{graphicx}
\usepackage[caption=false]{subfig} 
\usepackage{amsmath}
\usepackage{amssymb}

\newcommand{\um}{$\mu$m}

\newcommand{\MgII}{Mg\,{\small II}\,$\lambda$2800}

\newcommand{\hbeta}{H{$\beta$}}
\newcommand{\halpha}{H{$\alpha$}}

\newcommand{\Sersic}{S\'ersic}
\newcommand{\HST}	{\emph{HST}}%
\newcommand{\galfit}{\textsc{Galfit}}
\newcommand{\rr}    {https://youtu.be/dQw4w9WgXcQ?si=jQDYl1GPrmZN_1Ac}
\newcommand{\hrr}[2][black]{\href{\rr}{\color{#1}#2}}

\begin{document}
\title{Frequency and Abundance of Binary sUpermassive bLack holes from Optical Variability Surveys (FABULOVS): Hubble Space Telescope Imaging of Radial-Velocity Selected Binary Candidates}
\author[orcid=0000-0001-5769-0821, gname='Liam', sname='Nolan']{Liam Nolan}
\affiliation{Department of Astronomy, University of Illinois at Urbana-Champaign, Urbana, IL 61801, USA}
\email[show]{liamjn2@illinois.edu} 

\author[orcid=0000-0001-5105-2837, gname=Ming-Yang, sname=Zhuang]{Ming-Yang Zhuang} 
\affiliation{Department of Astronomy, University of Illinois at Urbana-Champaign, Urbana, IL 61801, USA}
\email{mingyang@illinois.edu} 

\author[orcid=0000-0003-0049-5210, gname=Xin, sname=Liu]{Xin Liu} 
\affiliation{Department of Astronomy, University of Illinois at Urbana-Champaign, Urbana, IL 61801, USA}
\affiliation{National Center for Supercomputing Applications, University of Illinois at Urbana-Champaign, Urbana, IL 61801, USA}
\email[show]{xinliuxl@illinois.edu}

\author[orcid=0000-0002-9932-1298, gname=Yu-Ching, sname=Chen]{Yu-Ching Chen}
\affiliation{Department of Physics and Astronomy, Johns Hopkins University, Baltimore, MD 21218, USA}
\affiliation{Department of Astronomy, University of Illinois at Urbana-Champaign, Urbana, IL 61801, USA}
\email{ycchen@jhu.edu}

\author[orcid=0009-0004-2898-6542, gname=Shreya, sname=Majumdar]{Shreya Majumdar} 
\affiliation{Department of Astronomy, University of Illinois at Urbana-Champaign, Urbana, IL 61801, USA}
\email{shreya.majumdar2002@gmail.com}

\author[orcid=0000-0002-1605-915X, gname=Junyao, sname=Li]{Junyao Li} 
\affiliation{Department of Astronomy, University of Illinois at Urbana-Champaign, Urbana, IL 61801, USA}
\email{junyaoli@illinois.edu}

\author[orcid=0000-0003-1659-7035, gname=Yue, sname=Shen]{Yue Shen} 
\affiliation{Department of Astronomy, University of Illinois at Urbana-Champaign, Urbana, IL 61801, USA}
\affiliation{National Center for Supercomputing Applications, University of Illinois at Urbana-Champaign, Urbana, IL 61801, USA}
\email{shenyue@illinois.edu}

\begin{abstract}
Sub-parsec (sub-pc) binary supermassive black holes (BSBHs) should be common from galaxy mergers, yet direct evidence has been elusive. We present HST/WFC3IR F160W imaging for a sample of 8 candidate sub-pc BSBHs at redshifts $z\sim$0.1--0.5, as well as cross-comparison with a sample of ordinary quasars with archival HST/WFC3 IR F160W images. These 8 candidate sub-pc BSBHs were identified from multi-epoch spectroscopic surveys of quasars (including both typical quasars and those with single-peaked velocity-offset broad lines). whose broad H$\beta$ lines are significantly offset (by $\gtrsim$ a few hundred km/s) from the systemic redshifts. We directly test the prediction that the host galaxies of BSBHs would have a higher fraction of disturbed morphologies and younger stellar bulges from recent interactions than those of control quasars. After careful subtraction of the central quasar light, our candidate BSBH hosts show a statistically undifferentiated distribution of host asymmetry, indicative of a similar fraction of recent mergers. While a significantly larger sample is needed to place this result on a much firmer statistical ground, it opens questions as to the timescale differences between galaxy merger and BSBH formation, or the efficacy of the radial-velocity-shift--based selection of sub-pc BSBH candidates.
  
\end{abstract}



\section{Introduction} \label{sec:intro}

LIGO has detected gravitational waves (GWs) from stellar-mass binary black hole mergers \citep{LIGO_observation_2016}, yet many GW sources are expected outside the LIGO frequency \citep{sesana_multi-band_2017, schutz_gravitational-wave_2018}. In particular, a low-frequency background detected by NANOGrav \citep{agazie_nanograv_2023} is predicted to be produced by the slow infall of Binary Supermassive Black Holes (BSBHs). A BSBH consists of two black holes, each with a mass of $\sim10^6$--$10^9$ M$_{\odot}$. BSBHs are expected to frequently form in galaxy mergers \citep{begelman_massive_1980, haehnelt_multiple_2002, volonteri_assembly_2003}, given that most massive galaxies harbor SMBHs \citep{kormendy_inward_1995, ferrarese_supermassive_2005}. The final coalescences should produce the loudest GW sirens in the universe \citep{thorne_gravitational-wave_1976, haehnelt_low-frequency_1994, vecchio_galaxy_1997, jaffe_gravitational_2003}, which will be the primary source of low-frequency GW experiments \citep{amaro-seoane_laser_2017, arzoumanian_nanograv_2018, sesana_testing_2018}. BSBHs are important for testing general relativity in the strong field regime and for the studies of galaxy evolution and cosmology \citep{centrella_black-hole_2010, merritt_dynamics_2013, colpi_massive_2014, berti_testing_2015}. 

While the formation of BSBHs seems inevitable \citep{begelman_massive_1980}, direct observational evidence has been elusive. No confirmed case is known at sub-pc scales, in stark contrast to theoretical expectations \citep{begelman_massive_1980, yu_evolution_2002}. Theory suggests that the orbital decay of BSBHs may slow down or stall at $\lesssim$pc scales \citep{begelman_massive_1980, milosavljevic_formation_2001, zier_binary_2001, yu_evolution_2002, vasiliev_new_2013, dvorkin_nightmare_2017, tamburello_supermassive_2017}, but the barrier may be overcome in gaseous environments \citep{gould_binary_2000, escala_role_2004, cuadra_massive_2009, lodato_black_2009, chapon_hydrodynamics_2013, rafikov_origin_2012, del_valle_supermassive_2015}, in triaxial or axisymmetric galaxies \citep{yu_evolution_2002, berczik_efficient_2006, preto_fast_2011, khan_supermassive_2013, khan_swift_2016, vasiliev_final-parsec_2015, gualandris_collisionless_2017, kelley_massive_2017}, and/or by interacting with a third supermassive black hole (SMBH) in hierarchical mergers \citep{valtonen_triple_1996, blaes_kozai_2002, hoffman_dynamics_2007, kulkarni_formation_2012, tanikawa_merger_2014, bonetti_post-newtonian_2018}. If a BSBH stalls at $\lesssim$pc scales, it will not emit GWs efficiently. This so-called ``Final Parsec Problem'' poses a significant uncertainty for the expected frequency of GW sources. Quantifying the occurrence rate of BSBHs below parsec scales is important to testing the theories of BSBH evolution to shed light on the rate of BSBH mergers.

Indirect approaches are necessary for identifying sub-pc BSBHs, whose typical separations are unresolvable even with VLBI (i.e., $\sim$mas) \citep{burke-spolaor_radio_2011}. At sub-centi-parsec scales, candidates have been proposed from quasi-periodic quasar light curves \citep[e.g.,][]{graham_possible_2015, liu_systematic_2016, zheng_sdss_2016, charisi_testing_2018,chen_candidate_2020,liao_discovery_2021,luo_systematic_2024}, most notably in OJ 287 \citep{valtonen_massive_2008}, but alternative scenarios remain possible (e.g., radio jet precession and/or false periodicity \citep{vaughan_false_2016}). Alternatively, sub-pc BSBHs may be identified from bulk broad-line radial velocity (RV) shifts as a function of time \citep{gaskell_quasars_1983, 
bogdanovic_modeling_2008, boroson_candidate_2009, shen_identifying_2010, mckernan_detection_2015, nguyen_emission_2016, pflueger_likelihood_2018}, in analogy to RV detections of exoplanets \citep{guo_constraining_2019}. Only one BH is assumed to be active, powering its own broad line region (BLR). The binary separation is much larger than the BLR size, such that the broad-line velocity traces the binary motion, yet small enough that the acceleration is detectable over the time baseline of typical observations \citep{runnoe_large_2015}. Previous work has mostly focused on a small sample of quasars with extremely broad, double peaks (velocity offsets $>$ a few thousand km s$^{-1}$) \citep{eracleous_completion_2003, boroson_candidate_2009, tsalmantza_systematic_2011}. Originally proposed as BSBHs where both BHs are active \citep{gaskell_evidence_1996}, these extremely wide double-peaked broad lines are now proved to be most likely rotation and relativistic effects in the accretion disks around single BHs (so-called ``disk emitters''; \citealt{eracleous_double-peaked_1999,strateva_double-peaked_2003,gezari_long-term_2007,lewis_long-term_2010}).

Unlike previous work, here we focus on a sample of strong BSBH candidates selected from ordinary quasars (\citealt{shen_constraining_2013}; see also: \citealt{ju_search_2013, wang_searching_2017}) and those with single-peaked, offset broad lines (\citealt{liu_constraining_2014}; see also: \citealt{eracleous_large_2012, runnoe_large_2015, runnoe_large_2017}). Using cross-correlation analysis, the broad \hbeta\ shifts have been measured in the largest extant sample of quasars with multi-epoch spectra \citep{shen_constraining_2013, liu_constraining_2014}. This dataset combines both repeated SDSS spectra for $\sim$700 ordinary quasars \citep{shen_constraining_2013} and dedicated, long-term follow-up observations for $\sim$50 kinematically offset quasars \citep{liu_constraining_2014}. Further third- and more-epoch spectra have been taken to test the binary hypothesis for the strongest candidates among this sample \citep{guo_constraining_2019}. In a BSBH with a constant acceleration, the velocity shifts are expected to be a few hundred km s$^{-1}$ in a few years, with no significant change in the broad emission line profile \citep{runnoe_large_2017, wang_searching_2017}. In contrast, BLR kinematics around single BHs (due to either long-term structural changes and/or short-term asymmetric reverberation of the BLRs in response to the varying continuum \citep{barth_lick_2015}) should exhibit stochastic accelerations and/or changes in the broad emission line profile. One must also consider the possibility of a GW recoil due to the anisotropic emission of GWs from the merger, in which the broad-line velocity shifts (and profiles) remain constant over human timescales.


These systematic, long-term spectroscopic monitoring studies have found 8 strong sub-pc BSBH candidates for further tests. The candidates show significant RV shifts in the broad \hbeta\ lines (corroborated by either broad \halpha\ or \MgII) over a few years (rest frame), yet with no significant change in the emission-line profile. The RV shifts can be explained by a 0.05--0.1 pc BSBH with an orbital period of $\sim$100 yr, assuming a mass ratio of $\sim$1 and a circular orbit \citep{guo_constraining_2019}. However, the spectroscopic monitoring studies represent only a first step toward finding BSBHs \citep{runnoe_large_2017}. Complementary tests are necessary, because the time baselines required to directly confirm the BSBH candidates are much longer than human lifetimes (e.g., at least a few centuries to cover several orbital cycles for BSBHs).


In this work, we present new high-resolution imaging of these 8 quasars as among the strongest known candidates of RV-selected sub-pc BSBHs obtained with the \textit{Hubble Space Telescope} Wide Field Camera 3 (\HST/WFC3). The observations represent the first dedicated imaging study of the host galaxies of quasars showing coherent broad-line velocity drifts. Combined with a carefully chosen control sample of ordinary quasars with existing archival HST images, this comparative study tests the merger hypothesis for the candidate BSBHs. This imaging can corroborate and help refine the spectroscopic technique for selecting BSBHs, potentially providing direct evidence for low-frequency GW sources for fundamental physics, such as those laid out by \citet{agazie_nanograv_2023}. We measure how frequently in our sample a sub-pc BSBH candidate is to be in a recently merged galaxy. More specifically, we quantify the fraction of disturbed host morphologies and interactions in the host galaxies, and compare such to the general active galactic nucleus (AGN) host galaxy population. Recent major mergers (with significant gas mass) should host a high fraction of young (i.e., high \Sersic\ index; \citealt{sersic_influence_1963}) bulges, disturbed host morphologies, and/or tidal features in the galactic outskirts. 

If the hosts of our candidate BSBHs display an elevated rate of merger indicators over that of the control sample, this both lends credence to the spectroscopic selection technique as well as the merger hypothesis. If however the host galaxies of candidate BSBHs are indistinguishable from those of the control sample, we must re-examine both for alternative formation channels and inadequacies. It is possible that there could be a significant difference in the merger timescale of SMBHs and their hosts, leading to merger features in hosts with a single SMBH, or fully relaxed hosts of BSBHs \citep[as suggested by][]{bardati_signatures_2024, izquierdo-villalba_properties_2023}. Minor mergers, unusual merger geometry, or extremely gas-rich host disks could also appear relatively relaxed while still retaining a BSBH. In any above case, the new \HST\ imaging helps to constrain the nature of quasars with kinematically offset broad emission lines by providing vital information about the hosts.

The paper is organized as follows. Section \ref{sec:data} describes how we developed the sample of interest (\S \ref{sec:soi}), how our \HST\ observations were performed (\S \ref{sec:obs}), how we selected our control sample of regular quasars (\S \ref{sec:cos}), how we brought our new and archival data into consistency for analysis (\S \ref{sec:red}), and how we characterized and modeled our point-spread functions (PSFs; \S \ref{sec:psf}). In Section \ref{sec:ana} we discuss our comparative analysis of the candidate BSBHs and control sample, both by modelling (\S \ref{sec:gal}), and other useful metrics (\S \ref{sec:met}). We show the results of this analysis alongside visual inspection in Section \ref{sec:res}, and discuss our interpretation of such in Section \ref{sec:dis}, along with proposed future work. We summarize and contextualize this work in Section \ref{sec:con}.

\section{Observations and Data Collection} \label{sec:data}

\subsection{Sample of Interest} \label{sec:soi}

Our targets are a sample of 8 Sloan Digital Sky Survey (SDSS) quasars at redshifts $0.1<z<0.5$. They represent currently the strongest RV-selected candidates known for hosting sub-pc BSBHs. They were selected from long-term, spectroscopic monitoring studies using the largest sample of quasars with multi-epoch spectroscopy \citep{shen_constraining_2013, liu_constraining_2014, guo_constraining_2019}. The parent quasar sample includes $\sim$2000 pairs of observations in total of which $\sim$700 pairs have good measurements (1 $\sigma$ error $\sim$40 km s$^{-1}$) of the velocity shifts between two epochs measured from cross-correlation analysis. Among SDSS DR7 spectroscopic quasars with two epoch spectra, 52 objects were identified with significant velocity shifts of a few hundred km s$^{-1}$ in broad emission lines. This shift was measured between two epochs (either from repeated SDSS observations \citep{shen_constraining_2013}, or by combining with new follow-up observations \citep{liu_constraining_2014}) separated by a few years (rest frame). From these, 16 candidate sub-pc BSBHs were further selected by demanding: (1) the measured velocity shift is caused by an overall shift in the bulk velocity, rather than variation in the broad line profiles; and (2) the velocity shifts independently measured from a second broad line (either broad \halpha\ or \MgII ) are consistent with those measured from broad \hbeta, which in combination greatly reduces false positives from transient blazar effects.

Further third- and fourth-epoch spectroscopy was conducted for the 16 candidates for continued RV tests \citep{guo_constraining_2019}. The long-term RV shifts suggests that 8 of the 16 quasars remain valid as strong BSBH candidates, which were observed in our \HST\ imaging program (see below). They show broad-line RV curves that are consistent with binary orbital motion without significant change in the broad line profiles. 
The RV shifts can be explained by a 0.05--0.1 pc BSBH with an orbital period of $\sim$100 yr, assuming a mass ratio of $\sim$1 and a circular orbit, although the parameters are not well constrained given the few epochs that sample only a small portion of the purported binary orbital cycle. No single-SMBH scenario easily mimics this constant acceleration. We note that, by construction, our sample is entirely composed of active SMBHs with broad lines, which excludes gas-poor mergers or inactive BSBHs - we probe only the subset of BSBHs which are actively accreting. Relevant details of the individual AGN hosts in our sample of interest (SoI) and control sample (as described in \S\ref{sec:con}) are listed in Table \ref{tab:ref}.

\begin{table*}
\centering
\caption{
Names and coordinates of AGN hosts from sample of interest (SoI)
and control sample.  $M_i$ is SDSS $i$-band magnitude; virial mass of the AGN central BH is derived from \hbeta.
}
\label{tab:ref}
\begin{tabular}{ ccccccc }
\hline
Name & Sample & z & RA & DEC & $M_i$ & Virial Mass [log10($M_{\odot}$)] \\
\hline\hline
J0847+3732 & SoI & 0.4534 & 131.8168 & 37.5383 & -24.262 & 8.238 \\
\hline
J0852+2004 & SoI & 0.4615 & 133.1542 & 20.0697 & -24.256 & 8.341 \\
\hline
J0928+6025 & SoI & 0.2959 & 142.1582 & 60.4225 & -24.403 & 8.999 \\
\hline
J1112+1813 & SoI & 0.1952 & 168.1288 & 18.2198 & -22.535 & 7.843 \\
\hline
J1229-0035 & SoI & 0.4498 & 187.2897 & -0.5917 & -24.003 & 8.668 \\
\hline
J1345+1144 & SoI & 0.1264 & 206.4521 & 11.7454 & -22.565 & 8.021 \\
\hline
J1410+3643 & SoI & 0.4495 & 212.5857 & 36.723 & -23.95 & 8.369 \\
\hline
J1537+0055 & SoI & 0.1365 & 234.2748 & 0.923 & -22.598 & 7.543 \\
\hline
J0035-1047 & Control & 0.7967 & 8.8203 & -10.792 & -24.773 & 8.348 \\
\hline
J0058-1026 & Control & 0.4898 & 14.6623 & -10.4476 & -24.542 & 8.973 \\
\hline
J0831+5034 & Control & 0.5308 & 127.9937 & 50.5789 & -23.74 & 8.211 \\
\hline
J0930+2402 & Control & 0.5592 & 142.6134 & 24.0424 & -23.967 & 9.258 \\
\hline
J1118+2336 & Control & 0.3814 & 169.5005 & 23.6143 & -23.484 & 9.23 \\
\hline
J1147-0146 & Control & 0.5531 & 176.9338 & -1.7671 & -24.37 & 8.146 \\
\hline
J1211+1438 & Control & 0.5804 & 182.8278 & 14.6362 & -24.705 & 8.339 \\
\hline
IBZF08023 & Control & 0.5724 & 317.3637 & -6.1709 & -24.507 & 9.058 \\
\end{tabular}
\end{table*}

\subsection{New HST/WFC3 Observations} \label{sec:obs}

In our \HST\ program (GO-15975; PI: X. Liu), we observed each candidate BSBH from UT 2020-05-12 13:22:40 to 2021-05-09 08:05:16 in rest-frame $\sim1$ \um\ with the F160W filter of WFC3. This waveband has the distinct advantage of reducing the impact of dust extinction and massive star forming regions on morphological measurements, and maximizing the host-quasar contrast. Our field-of-view (FoV) was $\sim400$ kpc at $z\sim0.5$ in the 512$\times$512 subarray -- sufficient to cover both host galaxy and close companions. We used the 4-step BOX dither pattern for cosmic-ray and hot-pixel rejection, and to better sample the point-spread function (PSF). Each target was observed for one orbit with an effective exposure time of 2055s reaching a typical 5$\sigma$ surface brightness limit of 25.93 mag/arcsec$^2$. Two dedicated PSF standard stars were observed with the same instrumental set up as with the science targets to help with PSF modeling (detailed below).

\subsection{Control Sample of Ordinary Quasars} \label{sec:cos}

\begin{figure}
    \includegraphics[width=\columnwidth]{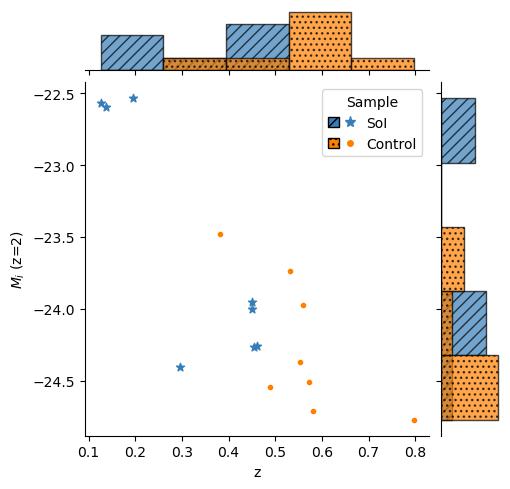}
    \caption{We selected our control sample of ordinary AGN (orange) to match to our sample of interest (blue) in SDSS i-band magnitude (or equivalent) and redshift (z)  as well as possible using KNN regression. When performing the Kolmogorov-Smirnov (KS) test on the distributions of each of these values, we get for  redshift a statistic of 0.875 (p-value 0.002), and for i mag a statistic of 0.5 (p-value 0.283). Despite KNN regression, significant cuts in sources appropriate for comparison (as described in \S\ref{sec:con}) makes better matching between the samples extremely difficult.}
    \label{fig:mat}
\end{figure}

In order to draw conclusions about how our sample compares to ordinary quasars, we developed a control sample from low-redshift quasars with existing archival \HST\ images. 
We compiled a list of all SDSS DR17 quasars which lie within the coverage of available \HST\ archives (up to $\sim2016$), a total of 2439. \hrr{We make this catalog available upon request}; the coverage check was done automatically by checking coordinate-coverage overlaps, and has not been thoroughly vetted for data quality. Regardless, it remains a useful point for generating a large list from which one can make cuts as appropriate for ones science case.
From this catalog, we perform a series of cuts to require existing WFC3 F160W data (181 quasars), a reasonable (150s) minimum exposure time (176 quasars), and a maximum separation of 0\farcs1 between SDSS and \HST\ centroids (meaning source centroid, not observation centroid) to ensure good cross-matching (94 quasars). We then use the \texttt{sklearn NearestNeighbors}\ implementation of K-nearest neighbors (KNN) regression to obtain the 20 quasars in the sample that are most similar in redshift and SDSS absolute $i$-band magnitude. We show the matching of our control sample to the sample of interest in Figure \ref{fig:mat}. We downloaded all suitable archival imaging in F160W for these 20 quasars. We conducted further cuts to this control sample due to data quality as detailed below, leaving 12. Finally, after reviewing the observing programs which produced the observations we wished to use for our control purposes, we discarded 4 which came from dedicated observations of post-starburst (PSB) hosts, as PSBs are generally thought to arise from merger events, which would undesirably weight our control away from "ordinary" AGN. This necessarily biases our sample in the opposite direction -- away from sources more likely to contain merger indicators. It is, of course, difficult to impossible to fully disentangle a sample of AGN from samples of merging galaxies, as the two have been known to be related for decades \citep[e.g.]{barnes_dynamics_1992}. We thus proceeded with a control sample of 8 AGN.

\subsection{\HST\ Data Reduction} \label{sec:red}

For both the sample of interest (SoI) and control sample, we combine the calibrated WFC3 images (\texttt{.flt}) of each source using \texttt{drizzlepac}. For our SoI, we have four images per source, dithered to produce a sub-pixel resolution of 0.06"/pixel after drizzling. In the process, cosmic rays and background are subtracted from the final images. In the control sample, 8 of twenty either lacked sufficient dithers to achieve the same resolution, had data quality issues preventing identical drizzling, or the source was too close to an edge or cosmic rays for meaningful extraction of features. It is essential for our purposes in comparing faint tidal features that the resolution between the SoI and control be identical to avoid systematics caused by sampling differences. Thus, we eliminated these sources, with 12 remaining in the control that were drizzled in the same fashion, which by eye improved sharpness.

\subsection{PSF Modeling: Construction and Uncertainties} \label{sec:psf}

\begin{figure}
    \includegraphics[width=\columnwidth]{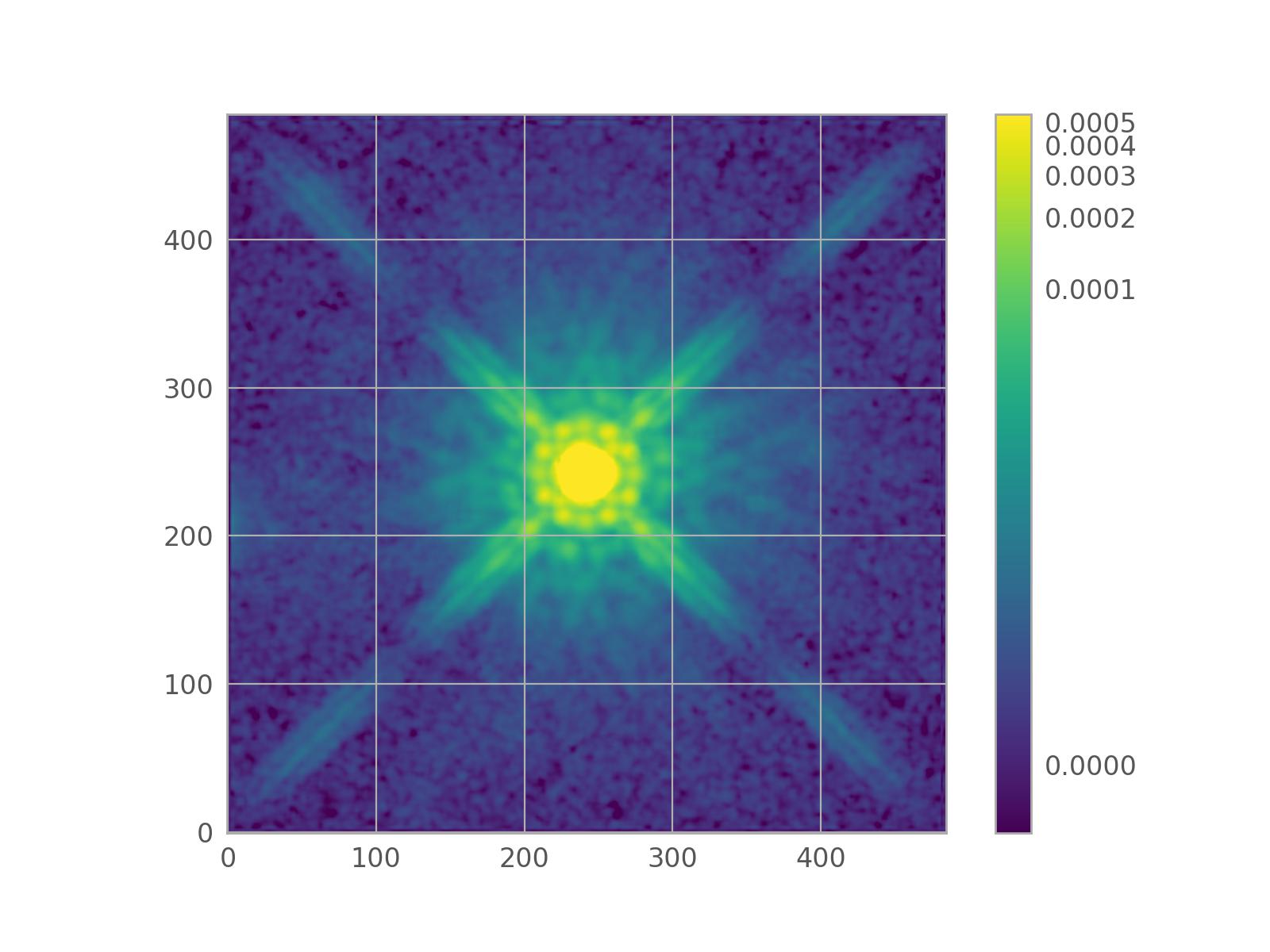}
    \caption{Effective PSF (ePSF) model as produced by \texttt{photutils EPSFBuilder} using our dedicated \HST\ PSF observations, as described in \S \ref{sec:psf}. Small mismatches between PSF observations and nuclear substructure produce small negative pixel values, but are not consequential for our analysis.}
    \label{fig:psf}
\end{figure}

We seek to characterize the features of the quasar hosts in both our SoI and control sample, but the AGN light itself dominates over the host-galaxy starlight in our \HST\ WFC3 images. We therefore require a model of the \HST\ point-spread function (PSF) in the images. We follow the recommended practice\footnote{See \href{https://www.stsci.edu/hst/instrumentation/focus-and-pointing/focus/tiny-tim-hst-psf-modeling}{STScI documentation on the subject.}} of developing an effective PSF \citep[ePSF, as described in][]{anderson_toward_2000} using \texttt{photutils EPSFBuilder} \citep{bradley_astropyphotutils_2023}. As part of our \HST\ observations, we observed two PSF stars with the same filters and dither patterns as our SoI, which form a robust basis for our ePSF. 
To augment these, we locate all bright sources in each image and extract them as small stamps, which we identify by inspection if they are both point-source-like and free of close neighbors.
We feed these bright sources and our dedicated PSF star exposures into \texttt{EPSFBuilder} with an oversampling factor of 3 and an iteration maximum of 10, each selected to balance accuracy and computational speed. We compare the performance of the ePSF with and without the bright sources in the SoI images, and find a clear improvement using solely the dedicated PSF exposures. The final SoI ePSF is shown in Figure \ref{fig:psf}.

As \HST\ PSFs vary on timescales of weeks to months \citep{kim_decomposition_2008}, our PSF model composed of sources across our SoI ought to characterize well the AGN therein, but likely declines in performance if applied to the older archival observations in the control sample. Indeed, the control observations are not designed to be close to each other in time, and so making any model composed of sources across the control observations would likely be erroneous. Therefore, we instead sought to develop a unique PSF model for each control observation wherein multiple clean point sources (free of close neighbors and artifacts, not overlapping with a known SDSS extended source) could be found. This unique PSF model would then be used in subtraction for such sources. There were two sources in our initial control sample which had suitably clean unique PSF models, but these later had to be removed from the sample for unrelated reasons. Therefore for sources in the control, we use our SoI-derived PSF model for consistency, 
accepting the likely but difficult to quantify loss in accuracy.

\section{Data Analysis} \label{sec:ana}

\subsection{Host Modeling} \label{sec:gal}


Recently merged galaxies are expected to have their gas driven inward, and therefore modestly elevated \Sersic\ indices \citep[][; though this relation is subject to greater uncertainty in AGN hosts]{conselice_evolution_2014}, and so modeling of our SoI and control sample as a superimposed central PSF and \Sersic\ profile is useful. In addition, as described further in Section \ref{sec:met}, the metrics by which we compare the rate of recent mergers in our samples are highly dependent on the asymmetric features of the host galaxies, which are to be found in the residuals of such a model. To accomplish said modeling, we use \galfit\ \citep{peng_detailed_2002, peng_detailed_2010}, a robust two-dimensional fitting algorithm further automated by our Python wrapper AFFOGATO\footnote{Available \href{https://github.com/LJNolan/BSBH-2022}{here.}}. Our procedure is as follows, for each source:

\begin{enumerate}
    \item Pull location information and excise a variable-size stamp based on the size of the source - from 12 to 24 arcsec$^2$.
    \item Use \texttt{photutils.segmentation}\ to mask other sources in the image which do not overlap with the target source, to reduce computational load of modelling and avoid skewing the best fit results. We perform two rounds of segmentation to ensure close but non-overlapping sources are caught.
    \item Pass masked stamp to \texttt{photutils.segmentation.SourceCatalog}\ to obtain estimates for target source (and any overlapping sources) of centroid position, Kron flux, half-light radius (approximated using \texttt{fluxfrac\_radius} as radius containing half the Kron flux), and ellipticity (inverse of elongation).
    \begin{itemize}
        \item Initial guesses for \Sersic\ index and position angle are set to 2 and $0^\circ$, respectively, as these parameters are thoroughly explored by \galfit\ and are difficult to estimate without fitting.
        \item We convert the estimated Kron flux to magnitude.
    \end{itemize}
    \item Pass parameter estimates to a \galfit\ input file, and create a constraint file to restrain parameters as follows, generally following \citet{zhuang_evolutionary_2023}:
    \begin{enumerate}
        \item centroid x,y: $\pm2$ pixels of initial position
        \item magnitude: $\pm5$ dex of initial estimate
        \item effective radius: 0.5 to 50 pixels
        \item \Sersic\ index: 0.3 to 8
        \item axis ratio: 0.1 to 1
    \end{enumerate}
    \begin{description}
        \item[Note] Due to degeneracy between the central PSF and the bulge of the host of the brightest quasars, there is a strong tendency for unconstrained models to wander and/or for the host \Sersic\ to become flatter and wider as the central PSF incorporates the bulge light \citep{zhuang_characterization_2024}. Thus we impose fairly wide but physically-motivated constraints (for an example, see \S\ref{app:gal}), and fits generally do not run up against these. If a parameter hits a constraint, we do not consider that model to have converged and reassess in fine-tuning.
    \end{description}
    \item Run \galfit\ using the stamp, mask, input, and constraint files created above.
    \item Review fits and make fine-tuning adjustments to recover convergence (if unconverged), make improvements, and/or check for degeneracy.
\end{enumerate}

\begin{figure*}
    \centering
    \subfloat[Automatic fit]{\includegraphics[width=\textwidth]{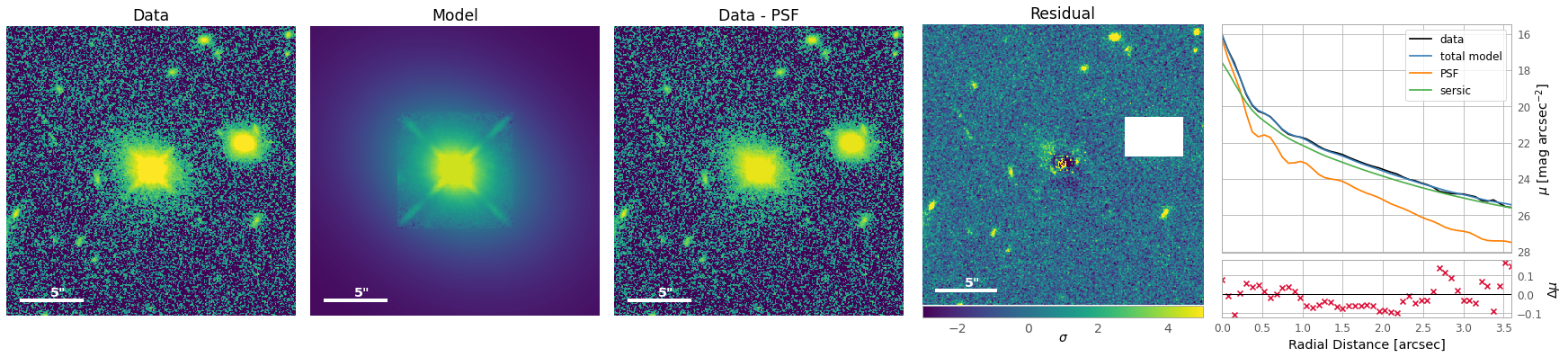}\label{fig:avf-a}}\\
    \subfloat[Manual fit]{\includegraphics[width=\textwidth]{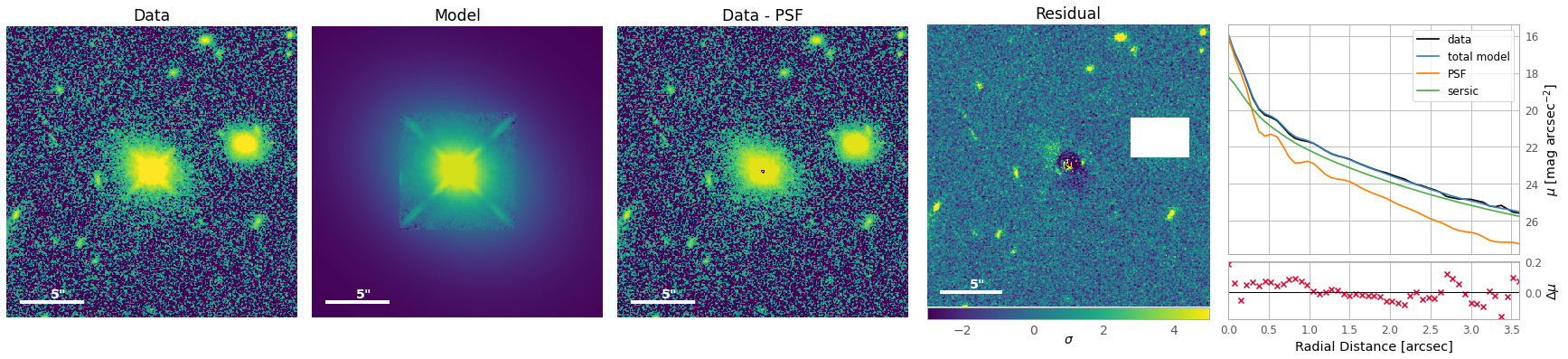}\label{fig:avf-b}}
    \caption{Comparison of an automatic fit result from AFFOGATO and a fine-tuned (manual) fit, both of source J1229-0035.}
    \label{fig:avf}
\end{figure*}

The differences between the automatic and fine-tuned fits are generally small (on the order of 0.01\%) in unrestricted parameters but can be significant in parameters which hit their bounding values. This is largely due to degeneracy between an unphysically concentrated host galaxy (\Sersic\ profile) and the central PSF, but manual adjustments are both symmetrical (and thus do not affect the asymmetrical host features we seek to investigate) and do not affect the appearance of the fit. It should also be noted that we do not make manual adjustments to the fitting of contaminants which hit boundary conditions, as we are not concerned with the fitting details of these contaminants and are rather concerned with efficient subtraction of their light from the central source's profile. We show an example in Figure \ref{fig:avf}. We report all fit information in Table \ref{tab:cat}, summarized in Section \ref{sec:res}, and examples of \galfit\ input and constraint files can be found in Appendix \ref{app:gal}.

\subsection{Other Metrics} \label{sec:met}

We measure asymmetry $A$ \citep{conselice_relationship_2003}, shape asymmetry $A_S$ \citep{pawlik_shape_2016}, and the Gini-M$_{20}$ merger statistic S(G, M$_{20}$) \citep{lotz_new_2004}, following the summary of these statistics by \citet{wilkinson_merger_2022}, using the \texttt{statmorph} \citep{rodriguez-gomez_optical_2019}\ package. In brief, $A$ is calculated by rotating a galaxy's profile by 180\textdegree\ and subtracting it from itself, $A_S$ is calculated in the same fashion with a flattened profile - all values set to 1, measuring only the deviation in shape rather than brightness. Finally, S(G, M$_{20}$) is a combination of two statistics: the Gini coefficient, and the second moment of the brightest 20\% of pixels, measuring the concentration and clumpiness of light in the galaxy, respectively. We reference the same threshold values ($A\geq0.35$, $A_S\geq0.4$, and S(G, M$_{20}$)$\geq0$) as \citet{wilkinson_merger_2022} when referencing a statistic as ``merger-indicating'', but rely primarily on comparisons between our SoI and control -- for these purposes, all three statistics are more indicative of merger as they increase. We perform these measurements of our SoI and control sample, using the drizzled images without model subtraction and with PSF subtraction. The data passed to \texttt{statmorph} is in the region defined by a \texttt{photutils} segmentation map to isolate the source from nearby contaminants in the same fashion as described in Section \ref{sec:gal}. All of these results, along with best-fit \Sersic\ indices, are reported in Table \ref{tab:met}. While asymmetry of the total model residual would be useful, the formalism for our statistics (and their implementation in \texttt{statmorph}) require positive total flux, and as expected, our residuals average approximately zero with large negative dips.  We instead rely on visual inspection for results derived from total residuals.

\section{Results} \label{sec:res}

In this section, we present the results of our analysis of the host galaxies of 8 candidate sub-pc binary SMBHs and a control sample of 8 ordinary quasars. We find no statistically significant differences in merger-related metrics (e.g., \Sersic\ index, asymmetry, Gini-M$_{20}$) between the two samples. However, visual inspection reveals potential asymmetrical features in some candidate BSBH hosts, suggesting that further investigation with a larger sample is needed.

\subsection{Host Galaxy Morphology} \label{sec:res:hgm}

We show the model parameters from \galfit\ for both our SoI and control sample in Table \ref{tab:cat}, and these models are reviewed visually in Figures \ref{fig:soi} \& \ref{fig:con} as discussed in \S\ \ref{sec:res:vir}. Although we matched the samples in SDSS i-band absolute magnitude (see Fig. Figure \ref{fig:mat}), our GALFIT results showed the candidate host+nucleus were on average 0.8 mag brighter/fainter than the control in F160W. This slight discrepancy could be due to differences in spectral energy distribution or sample matching. The effective radii ($R_{eff}$) of the SoI is also slightly larger on average than the control sample, consistent with the skew from the unusually large J1112+1813 (Fig. \ref{fig:soi4}) and J1345+1144 (Fig. \ref{fig:soi6}). There is no notable difference in the spread of \Sersic\ indices, as shown in Figure \ref{fig:sts}.

\begin{table*}
\centering
\caption{
\galfit\ model parameters and associated data, for the SoI and control
 sample.  Values above the following thresholds indicate mergers:
 $A\geq0.35$, $A_S\geq0.4$, S(G, M$_{20}$)$\geq0$ 
 \citep{wilkinson_merger_2022}.
}
\label{tab:cat}
\begin{tabular}{ ccccccccc }
\hline
Name & Sample & Fit $\chi^{2}/\nu$ & Component & Mag & $R_{eff} [pix]$ & Sérsic index & Axis Ratio & Pos. Angle \\
\hline\hline
J0847+3732 & SoI & 1.76 & sersic & 21.3 & 12.7 & 4.04 & 0.68 & 34.3 \\
 &  &  & psf & 19.9 &  &  &  &  \\
 &  &  & contaminant & 22.7 & 13.9 & 1.86 & 0.33 & 89.5 \\
 &  &  & contaminant & 22.0 & 12.2 & 1.63 & 0.62 & -58.9 \\
\hline
J0852+2004 & SoI & 1.37 & sersic & 20.5 & 9.6 & 2.99 & 0.83 & 38.5 \\
 &  &  & psf & 20.9 &  &  &  &  \\
\hline
J0928+6025 & SoI & 4.09 & sersic & 19.3 & 11.0 & 4.81 & 0.95 & -9.8 \\
 &  &  & psf & 19.0 &  &  &  &  \\
 &  &  & contaminant & 20.4 & 9.6 & 4.96 & 0.96 & 17.1 \\
\hline
J1112+1813 & SoI & 1.7 & sersic & 18.6 & 29.5 & 3.2 & 0.35 & 15.8 \\
 &  &  & psf & 19.9 &  &  &  &  \\
\hline
J1229-0035 & SoI & 1.88 & sersic & 20.9 & 8.8 & 4.2 & 0.77 & 56.8 \\
 &  &  & psf & 20.3 &  &  &  &  \\
\hline
J1345+1144 & SoI & 2.03 & sersic & 18.4 & 34.1 & 2.8 & 0.96 & -78.4 \\
 &  &  & psf & 23.8 &  &  &  &  \\
 &  &  & contaminant & 22.0 & 18.5 & 2.51 & 0.79 & 24.8 \\
 &  &  & contaminant & 22.2 & 20.6 & 1.66 & 0.72 & -62.1 \\
\hline
J1410+3643 & SoI & 1.58 & sersic & 20.9 & 15.3 & 2.21 & 0.9 & 36.0 \\
 &  &  & psf & 26.0 &  &  &  &  \\
 &  &  & contaminant & 23.7 & 5.0 & 3.82 & 0.68 & -44.7 \\
\hline
J1537+0055 & SoI & 2.9 & sersic & 18.0 & 12.1 & 1.82 & 0.97 & 21.7 \\
 &  &  & psf & 20.3 &  &  &  &  \\
\hline
003516.88-104731.2 & Control & 1.46 & sersic & 22.4 & 7.9 & 1.46 & 0.74 & 64.7 \\
 &  &  & psf & 21.5 &  &  &  &  \\
\hline
005838.95-102651.3 & Control & 1.46 & sersic & 20.8 & 7.8 & 3.07 & 0.33 & -33.8 \\
 &  &  & psf & 20.6 &  &  &  &  \\
\hline
083158.49+503444.1 & Control & 1.68 & sersic & 21.1 & 11.4 & 2.53 & 0.5 & 49.9 \\
 &  &  & psf & 21.3 &  &  &  &  \\
\hline
093027.21+240232.7 & Control & 2.64 & sersic & 20.4 & 16.6 & 2.58 & 0.48 & -79.5 \\
 &  &  & psf & 21.0 &  &  &  &  \\
\hline
111800.12+233651.5 & Control & 2.18 & sersic & 20.2 & 15.4 & 4.73 & 0.83 & -89.5 \\
 &  &  & psf & 20.9 &  &  &  &  \\
 &  &  & contaminant & 23.2 & 11.7 & 3.05 & 0.9 & 13.4 \\
\hline
114744.11-014601.5 & Control & 1.47 & sersic & 22.6 & 10.9 & 1.06 & 0.59 & 20.9 \\
 &  &  & psf & 20.9 &  &  &  &  \\
 &  &  & contaminant & 24.3 & 10.0 & 0.37 & 0.44 & 45.3 \\
\hline
121118.66+143810.4 & Control & 1.27 & sersic & 21.9 & 11.4 & 3.75 & 0.55 & 72.9 \\
 &  &  & psf & 21.7 &  &  &  &  \\
\hline
210927.29-061015.1 & Control & 2.15 & sersic & 20.9 & 11.4 & 3.94 & 0.77 & 53.1 \\
 &  &  & psf & 21.6 &  &  &  &  \\
\end{tabular}
\end{table*}

\subsection{Merger-Related Metrics} \label{sec:res:mrm}

\begin{figure*}
    \centering
    \subfloat[]{\includegraphics[width=0.49\textwidth]{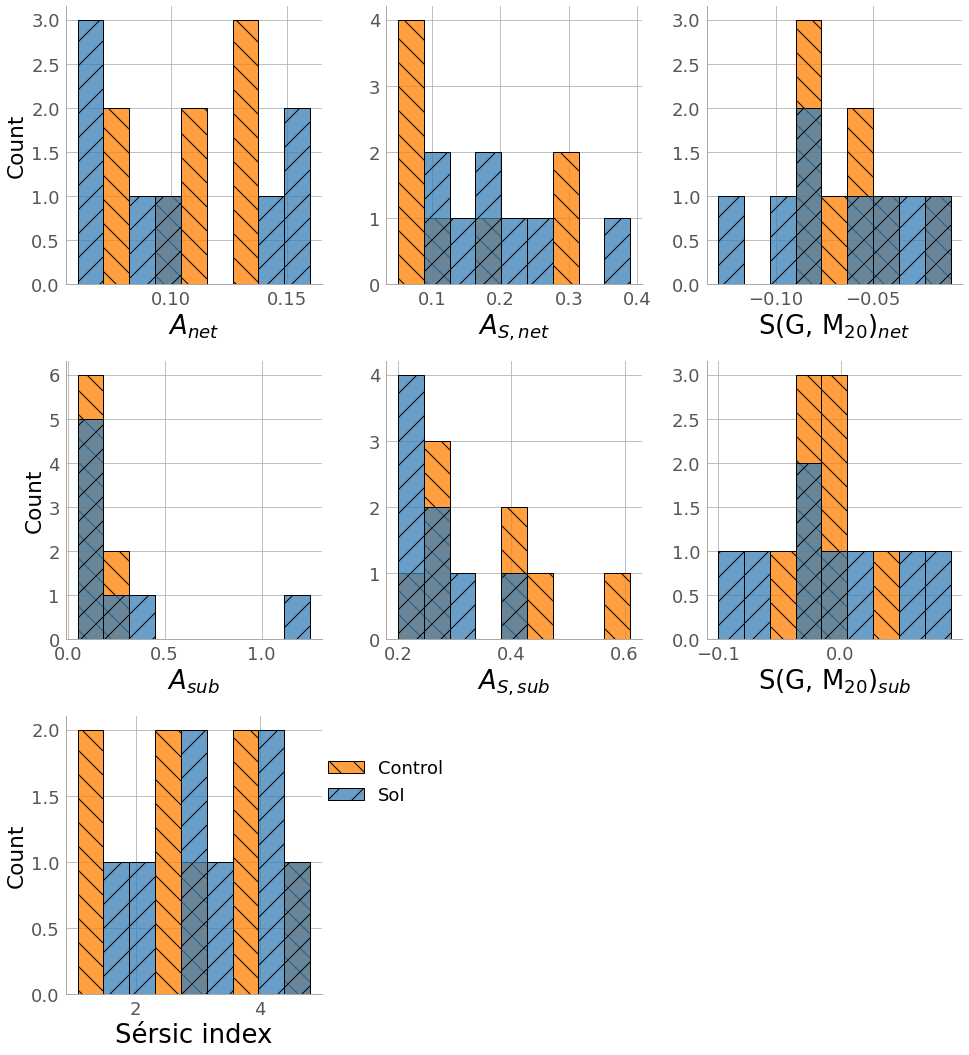}\label{fig:sth}}
    \subfloat[]{\includegraphics[width=0.49\textwidth]{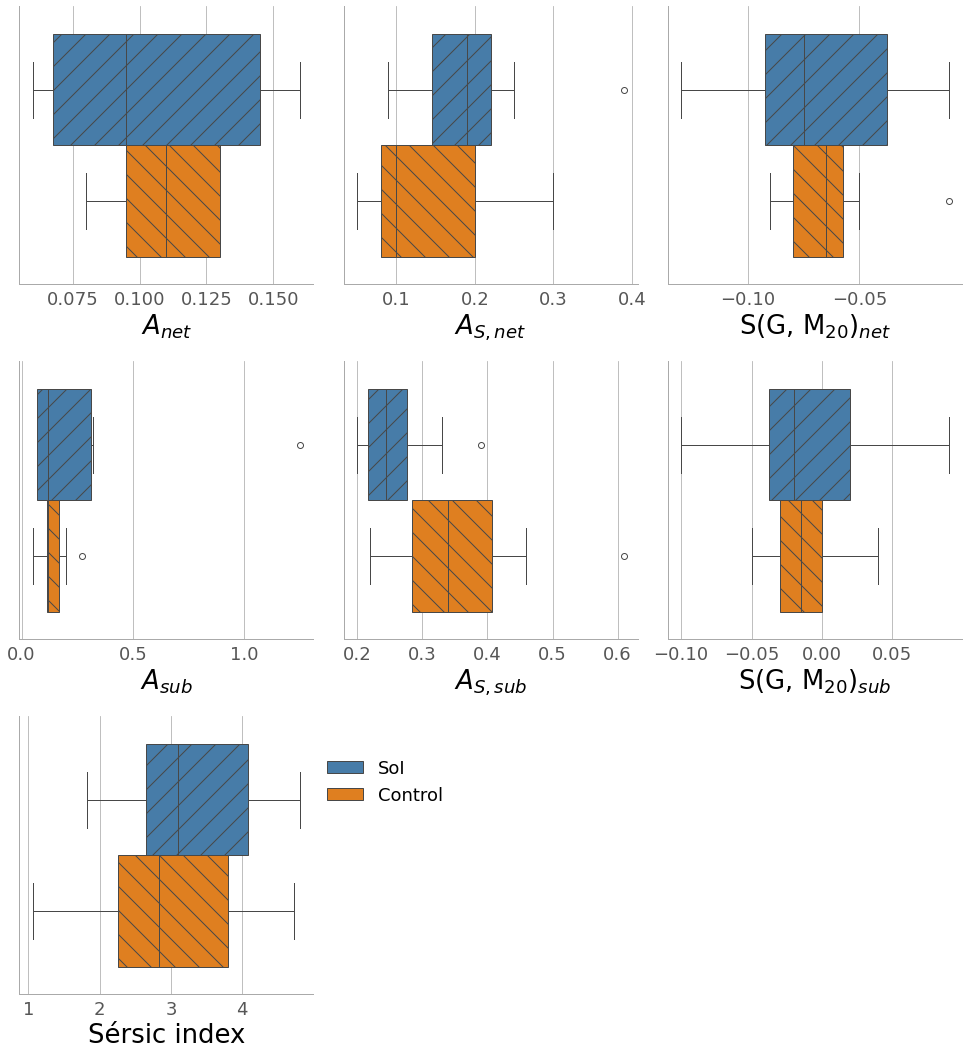}\label{fig:stb}}
    \caption{Merger-related statistics for our SoI (blue) and control (orange) samples, presented as both histograms and boxplots. We note difference between our samples only in PSF-subtracted shape asymmetry $A_{S, sub}$, but no differences are statistically significant (see \S\ref{sec:res:mrm}).}
    \label{fig:sts}
\end{figure*}

As described in \S\ \ref{sec:met}, we use the \Sersic\ index, asymmetry, shape asymmetry, and Gini-M$_{20}$ coefficient to both suggest merger status of individual galaxies in our samples, and analyze the relative merger rate between our samples. These statistics are listed in Table \ref{tab:met} and plotted in Figure \ref{fig:sts}. Our SoI shows an elevated average \Sersic\ index, $A_{S,net}$, $A_{sub}$, and S(G, M$_{20}$)$_{sub}$, and a depressed average $A_{net}$, S(G, M$_{20}$)$_{net}$, and $A_{S, sub}$. However, none of these differences are greater than 1$\sigma$. Higher \Sersic\ indices are modestly indicative of greater central concentration, which would be expected in recently merged hosts \citep{aceves_sersic_2006}, but again due to \Sersic-PSF degeneracy noted above (as well as uncertainty cited by \citet{conselice_evolution_2014} in \Sersic\-merger relations in AGN hosts), we do not adopt a \Sersic\ index criterion to indicate a specific galaxy is a merger. None of our full-science-image ($net$) statistics indicate mergers, in the SoI or control. This is unsurprising, as the AGN PSF is symmetrical and dominates the image. After PSF subtraction ($sub$), asymmetry indicates no mergers; shape asymmetry indicates ($A_S\geq0.4$) 1 out of 8 galaxies in the SoI are mergers, and 3 out of 8 in the control are mergers; Gini-M20 indicates (S(G, M$_20$)$\geq0$) 3 out of 8 galaxies in the SoI are mergers, and 3 out of 8 in the control are mergers \citep{wilkinson_merger_2022}. These statistics do not indicate an elevated rate of recent mergers among our candidate binary AGN hosts.

To determine whether our merger statistics give any indication of our SoI and control being drawn from distinct underlying populations (e.g., with and without BSBHs), we perform the K-S test and Mann-Whitney U test \citep{mann_test_1947}, and report the relevant statistics in Table \ref{tab:mst}. PSF-subtracted shape asymmetry has the lowest p-value of 0.087, but still lies above a threshold of 0.05. These tests therefore do not indicate our samples come from distinct populations -- nor, of course, do we prove our samples come from the same underlying population (especially due to small sample size), we simply cannot draw a conclusion in either direction.

\begin{table*}
\centering
\caption{
Merger statistics for the SoI and control sample; $A_{net}$, $A_{S, net}$, and S(G, M$_{20}$)$_{net}$ are the asymmetry, shape asymmetry, and Gini-M20 metrics as described in Section \ref{sec:met} for the full science image (i.e. AGN + host), and those with the -sub prefix are for the science image minus the central PSF model (i.e. just the host, with AGN subtracted). Values above the thresholds $A\geq0.35$, $A_S\geq0.4$, and S(G, M$_{20}$)$\geq0$ indicate a recent merger \citep{wilkinson_merger_2022}.}
\label{tab:met}
\begin{tabular}{ cccccccccc }
\hline
Name & Sample & Fit $\chi^{2}/\nu$ & \Sersic & $A_{net}$ & $A_{S, net}$ & S(G, M$_{20}$)$_{net}$ & $A_{sub}$ & $A_{S, sub}$ & S(G, M$_{20}$)$_{sub}$ \\
\hline\hline
J0847+3732 & SoI & 1.762693 & 4.0442 & 0.1578 & 0.095 & -0.0878 & 0.3169 & 0.262 & 0.0522 \\
\hline
J0852+2004 & SoI & 1.368682 & 2.9893 & 0.0902 & 0.1825 & -0.0065 & 0.1398 & 0.2163 & -0.0133 \\
\hline
J0928+6025 & SoI & 4.094249 & 4.81 & 0.1586 & 0.0926 & -0.1317 & 0.3107 & 0.2282 & 0.0113 \\
\hline
J1112+1813 & SoI & 1.697762 & 3.2039 & 0.0553 & 0.2472 & -0.0387 & 0.0714 & 0.255 & -0.0292 \\
\hline
J1229-0035 & SoI & 1.876276 & 4.2035 & 0.1392 & 0.1573 & -0.0863 & 1.2527 & 0.3308 & 0.0852 \\
\hline
J1345+1144 & SoI & 2.032509 & 2.7977 & 0.0645 & 0.3923 & -0.0311 & 0.0733 & 0.3927 & -0.0276 \\
\hline
J1410+3643 & SoI & 1.582821 & 2.2105 & 0.0665 & 0.2027 & -0.1019 & 0.0727 & 0.203 & -0.1035 \\
\hline
J1537+0055 & SoI & 2.89794 & 1.8182 & 0.1017 & 0.2137 & -0.0575 & 0.0988 & 0.2147 & -0.0589 \\
\hline
003516.88-104731.2 & Control & 1.460576 & 1.4558 & 0.1324 & 0.0828 & -0.0929 & 0.2006 & 0.3906 & -0.0063 \\
\hline
005838.95-102651.3 & Control & 1.455108 & 3.0701 & 0.1116 & 0.0787 & -0.0618 & 0.1229 & 0.2209 & 0.0037 \\
\hline
083158.49+503444.1 & Control & 1.679907 & 2.5324 & 0.1014 & 0.2948 & -0.0752 & 0.1635 & 0.6111 & -0.0299 \\
\hline
093027.21+240232.7 & Control & 2.644243 & 2.5824 & 0.1296 & 0.3018 & -0.0053 & 0.1211 & 0.3863 & -0.0536 \\
\hline
111800.12+233651.5 & Control & 2.177344 & 4.7269 & 0.085 & 0.0496 & -0.0594 & 0.1219 & 0.2882 & -0.0253 \\
\hline
114744.11-014601.5 & Control & 1.468846 & 1.0566 & 0.1314 & 0.1659 & -0.0682 & 0.2705 & 0.4579 & 0.037 \\
\hline
121118.66+143810.4 & Control & 1.274343 & 3.7522 & 0.1096 & 0.1225 & -0.0762 & 0.0971 & 0.2883 & 0.0013 \\
\hline
210927.29-061015.1 & Control & 2.151586 & 3.9386 & 0.0791 & 0.0753 & -0.0544 & 0.0518 & 0.2667 & -0.0214 \\
\end{tabular}
\end{table*}

\begin{table*}
\centering
\caption{
Merger statistic averages for the SoI and control sample, followed by the
difference (Control - SoI).
}
\label{tab:avg}
\begin{tabular}{ cccccccc }
\hline
Sample & \Sersic & $A_{net}$ & $A_{S, net}$ & S(G, M$_{20}$)$_{net}$ & $A_{sub}$ & $A_{S, sub}$ & S(G, M$_{20}$)$_{sub}$ \\
\hline\hline
SoI & 3.259662 & 0.104225 & 0.197912 & -0.067688 & 0.292037 & 0.262838 & -0.010475 \\
\hline
Control & 2.889375 & 0.110012 & 0.146425 & -0.061675 & 0.143675 & 0.36375 & -0.011813 \\
\hline
$\Delta_{c-s}$ & -0.370288 & 0.005788 & -0.051487 & 0.006013 & -0.148362 & 0.100913 & -0.001338 \\
\end{tabular}
\end{table*}

\begin{table*}
\centering
\caption{
Statistics ($S$) and p-values ($p$) from the Kolmogorov-Smirnov ($KS$) and Mann-Whitney U ($MWU$) tests run on merger statistics.  A p-value of $\leq$ 0.05 indicates our SoI and control come from different underlying populations.
}
\label{tab:mst}
\begin{tabular}{ cccccccc }
\hline
 & \Sersic & $A_{net}$ & $A_{S, net}$ & S(G, M$_{20}$)$_{net}$ & $A_{sub}$ & $A_{S, sub}$ & S(G, M$_{20}$)$_{sub}$ \\
\hline\hline
$S_{KS}$ & 0.25 & 0.375 & 0.5 & 0.375 & 0.375 & 0.625 & 0.25 \\
\hline
$p_{KS}$ & 0.98 & 0.66 & 0.283 & 0.66 & 0.66 & 0.087 & 0.98 \\
\hline
$S_{MWU}$ & 39.0 & 29.0 & 45.0 & 29.0 & 34.0 & 13.0 & 31.0 \\
\hline
$p_{MWU}$ & 0.505 & 0.798 & 0.195 & 0.798 & 0.878 & 0.05 & 0.959 \\
\end{tabular}
\end{table*}

\subsection{Visual Inspection of Residuals} \label{sec:res:vir}

We present individual analysis of the residuals of our two samples in Table \ref{tab:vis}. In summary, we note extended asymmetrical structure underneath the AGN PSF and host \Sersic\ profile in the SoI in Figures \ref{fig:soi2}, \ref{fig:soi3}, \ref{fig:soi5}, \ref{fig:soi6}, and \ref{fig:soi8}. We omit from this list asymmetrical features we are not confident come from underlying host features, and could instead be due to contaminants.  We do not note any features with this confidence in the control sample, but direct the reader to Figures \ref{fig:con3}, \ref{fig:con25}, \ref{fig:con40}, and \ref{fig:con85} as more tentatively disturbed candidates. While it is certainly possible to interpret this as an elevated rate of large-scale asymmetrical features in our SoI, some skepticism is warranted due to our PSF being more finely tuned to the SoI than the control; however, errors are not significantly higher among the control than the SoI, even in the central PSF-dominated regions, and additional re-scaling and re-stretching of the control residuals do not reveal ``washed-out'' features. We do not see any double nuclei at HST resolution, consistent with the sub-pc scale of the putative binaries.

\begin{table*}
\centering
\caption{
Visual analysis results of residuals, as seen in Figures \ref{fig:soi} \& \ref{fig:con}.
}
\label{tab:vis}
\begin{tabular}{p{0.2\textwidth}p{0.65\textwidth}}
\hline
Object & Notes \\
\hline\hline
\hyperref[fig:soi1]{J0847+3732} & No clear underlying features, slight asymmetry in PSF wings, could be related to close contaminant sources. \\
\hline
\hyperref[fig:soi2]{J0852+2004} & Lopsided wings extending to the right and bottom-left, approximately 1'' ($\sim5$kpc). \\
\hline
\hyperref[fig:soi3]{J0928+6025} & Roughly half (left) overbright, though some may be due to modeled contaminant to bottom-right. \\
\hline
\hyperref[fig:soi4]{J1112+1813} & Asymmetrical extended top-to-bottom feature, though source's elongation implies this is likely a closer to edge-on than face-on galaxy, and may be an effect thereof. \\
\hline
\hyperref[fig:soi5]{J1229+0035} & Small ($\lesssim1''$) asymmetrical wing on left. \\
\hline
\hyperref[fig:soi6]{J1345+1144} & Clear underlying spiral features, which extend into a tidal arm at top-right. \\
\hline
\hyperref[fig:soi7]{J1410+3643} & Bright ring around central AGN, appears similar to tight spiral arms or a sharply cut-off bulge. \\
\hline
\hyperref[fig:soi8]{J1537+0055} & Extremely clear underlying spiral, extending into a tidal arm to top-right -- this could be related to nearby contaminant but seems continuous with underlying structure. \\
\hline
\hyperref[fig:con2]{003516.88-104731.2} & Faint spiral arms. \\
\hline
\hyperref[fig:con3]{005838.95-102651.3} & Modest asymmetry is coincident with PSF arms and primary axis of galaxy. \\
\hline
\hyperref[fig:con14]{083158.49+503444.1} & Asymmetry of underlying features cannot be discerned due to large number of central contaminants or bright regions. \\
\hline
\hyperref[fig:con25]{093027.21+240232.7} & Small asymmetrical bright region on right. \\
\hline
\hyperref[fig:con40]{111800.12+233651.5} & Asymmetrical ring-like isophote coincident with point-like contaminant. \\
\hline
\hyperref[fig:con44]{114744.11-014601.5} & Small bright asymmetrical region to the top-right, aligned with modeled contaminant. \\
\hline
\hyperref[fig:con48]{121118.66+143810.4} & No visible extended features. \\
\hline
\hyperref[fig:con85]{210927.29-061015.1} & Modestly asymmetrical small arms, containing a bright point-like region. \\
\end{tabular}
\end{table*}

\begin{figure*}
    \centering
    \subfloat[J0847+3732]{\includegraphics[width=0.95\textwidth]{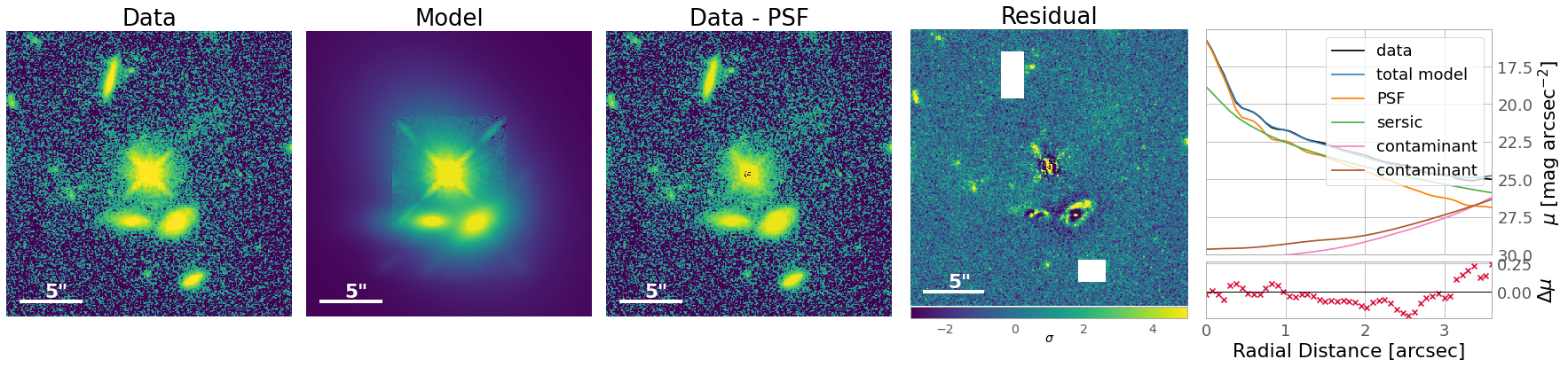}\label{fig:soi1}}\\
    \subfloat[J0852+2004]{\includegraphics[width=0.95\textwidth]{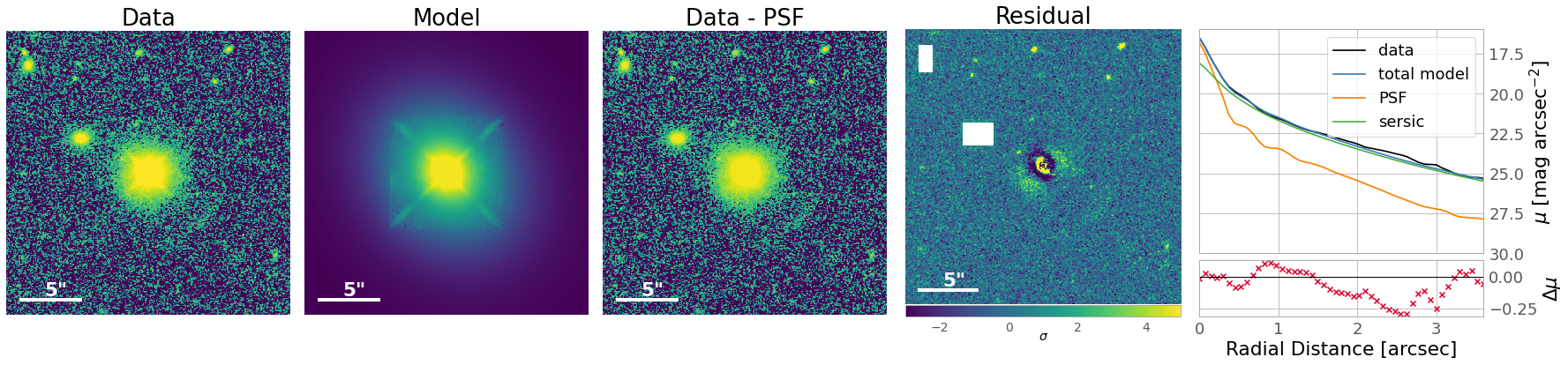}\label{fig:soi2}}\\
    \subfloat[J0928+6025]{\includegraphics[width=0.95\textwidth]{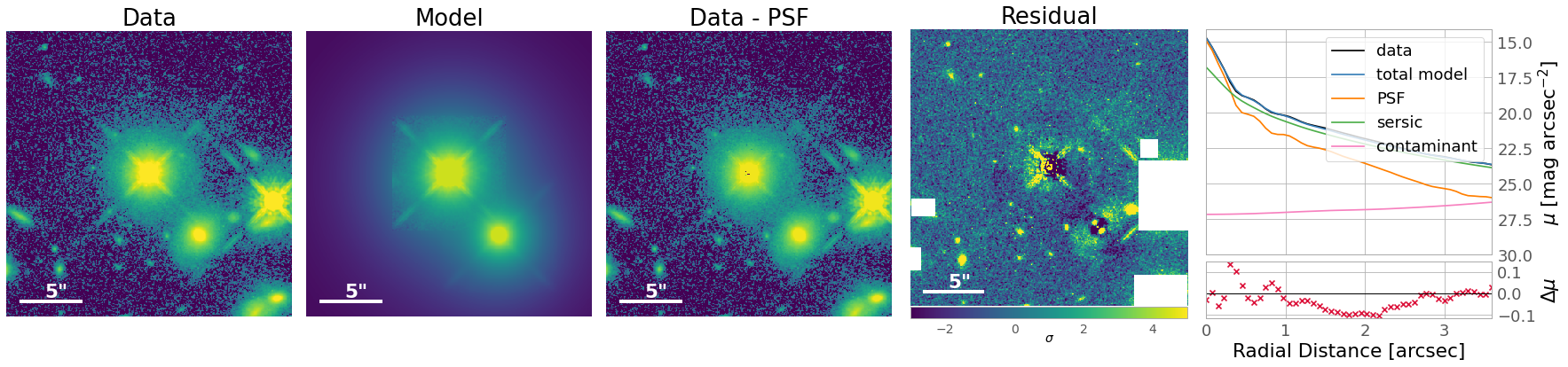}\label{fig:soi3}}\\
    \subfloat[J1112+1813]{\includegraphics[width=0.95\textwidth]{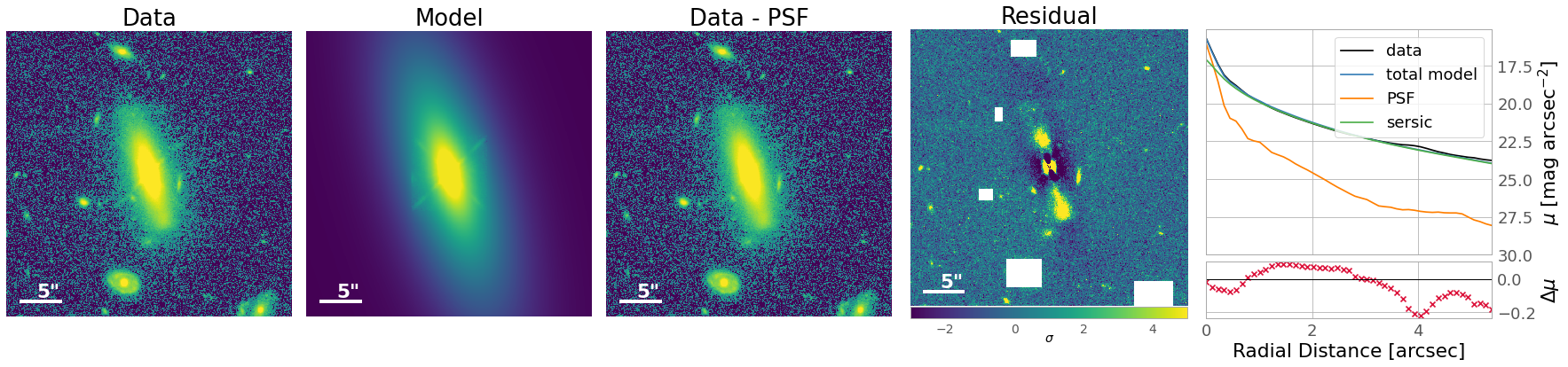}\label{fig:soi4}}
    \caption{}
\end{figure*}
\begin{figure*}\ContinuedFloat
    \subfloat[J1229-0035]{\includegraphics[width=0.95\textwidth]{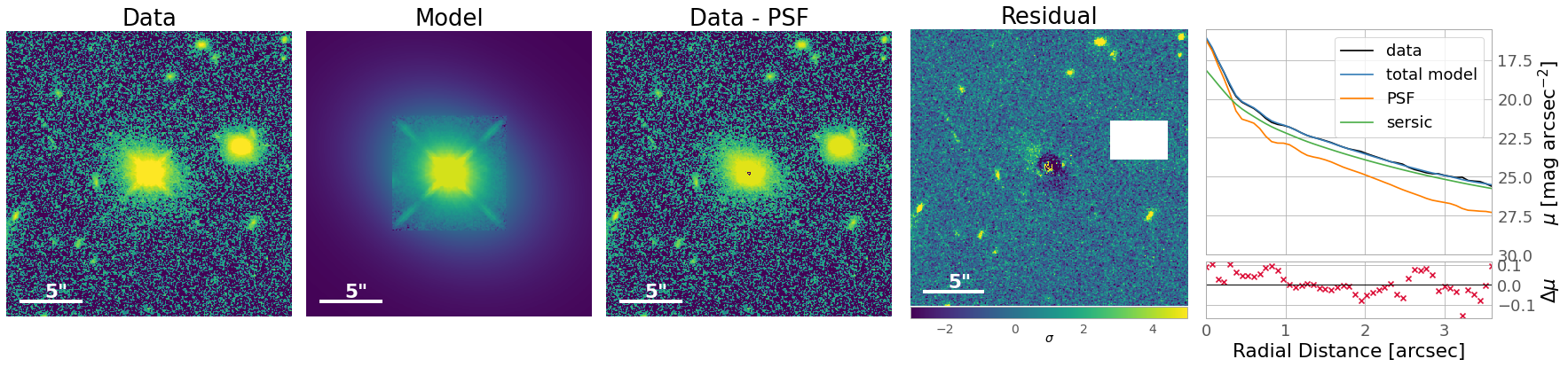}\label{fig:soi5}}\\
    \subfloat[J1345+1144]{\includegraphics[width=0.95\textwidth]{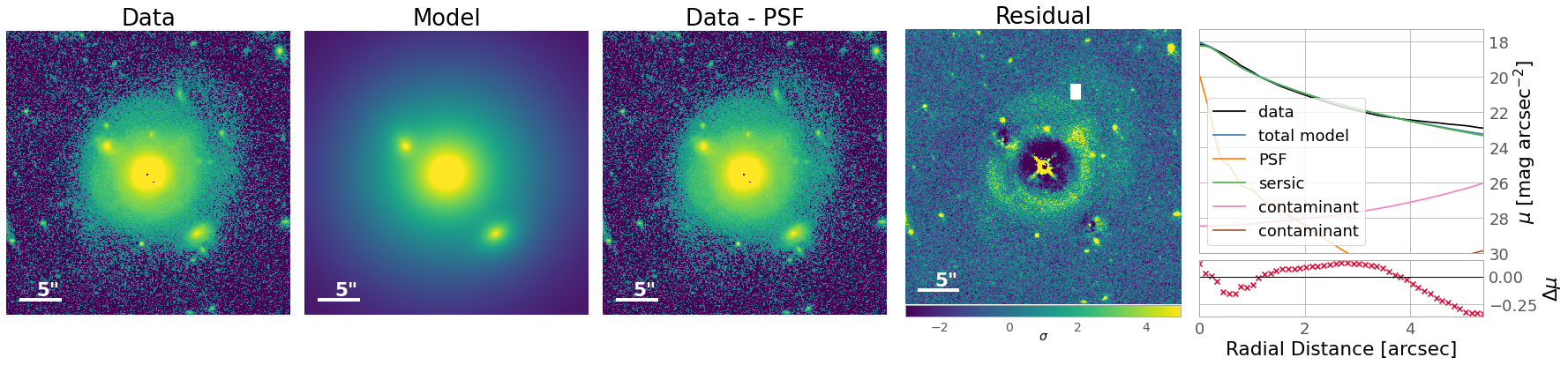}\label{fig:soi6}}\\
    \subfloat[J1410+3643]{\includegraphics[width=0.95\textwidth]{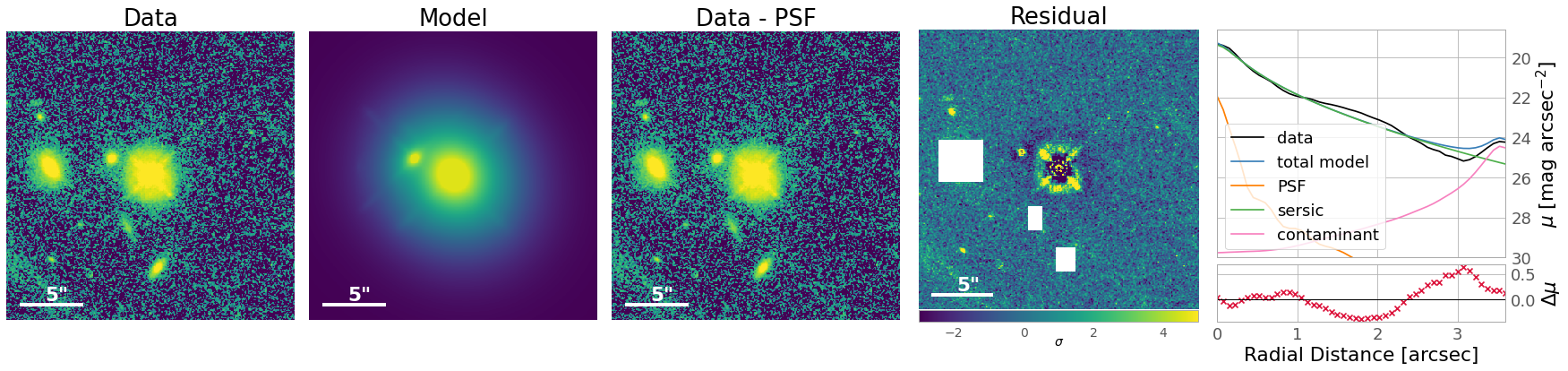}\label{fig:soi7}}\\
    \subfloat[J1537+0055]{\includegraphics[width=0.95\textwidth]{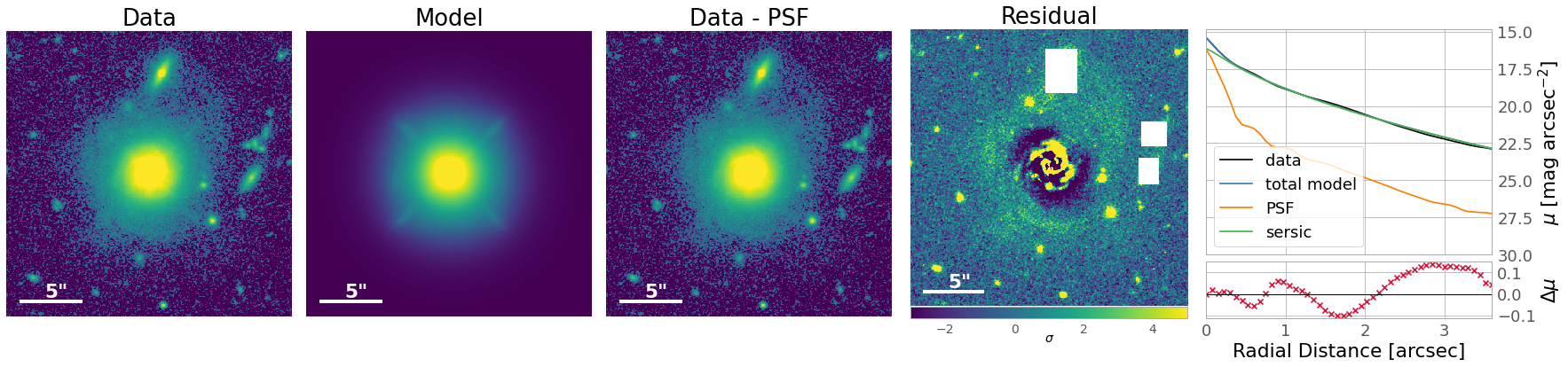}\label{fig:soi8}}
    \caption{Primary fitting results of our sample of interest. From left to right: \textit{Data:} drizzled \HST\ data cutout centered on target, log stretch; \textit{Model:} \galfit\ model, log stretch; \textit{Data-PSF:} \HST\ data with PSF component of \galfit\ model subtracted, intended to show the host galaxy without central AGN, log stretched; \textit{Residual:} difference image of \HST\ data and \galfit\ model, linear scale to \HST\ image $\sigma$ (shown), with white boxes indicating masked regions not considered by the model (usually major contaminants); Final panel: radial profile of central region, centered on AGN PSF, 'contaminant' referring to any \Sersic\ component used to model nearby non-target sources.}
    \label{fig:soi}
\end{figure*}
\begin{figure*}
    \subfloat[003516.88-104731.2]{\includegraphics[width=0.95\textwidth]{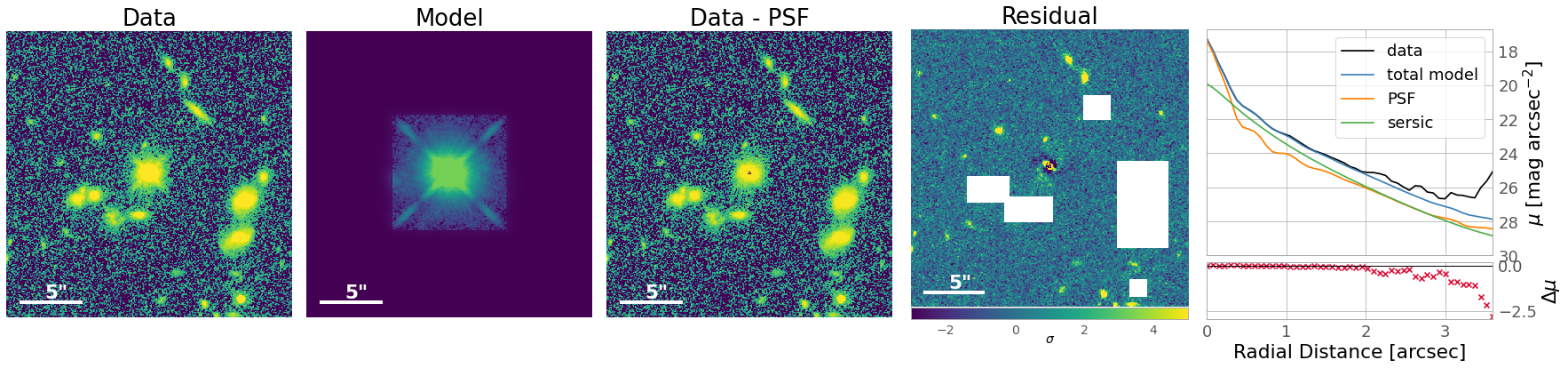}\label{fig:con2}}\\
    \subfloat[005838.95-102651.3]{\includegraphics[width=0.95\textwidth]{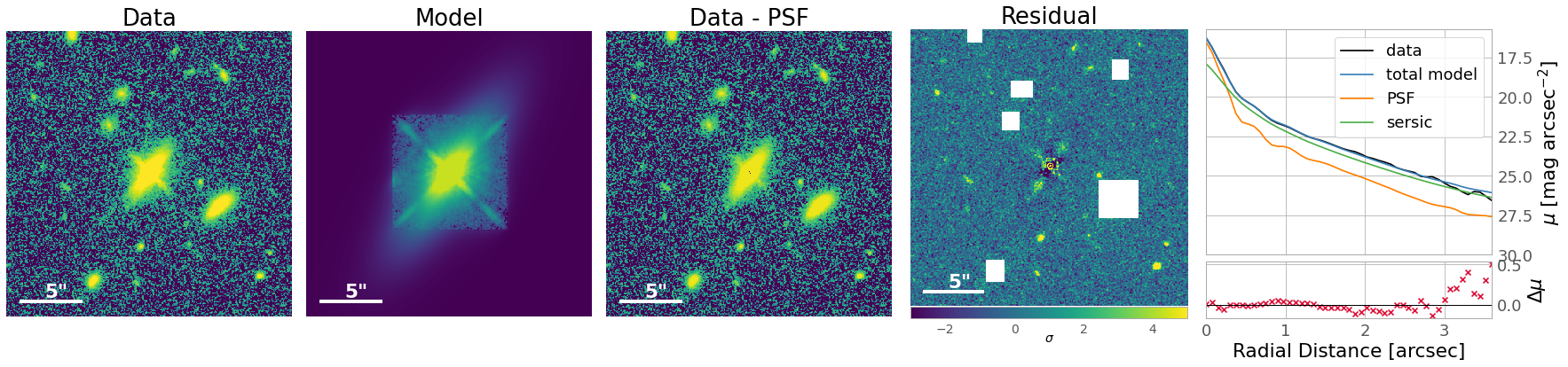}\label{fig:con3}}\\
    \subfloat[083158.49+503444.1]{\includegraphics[width=0.95\textwidth]{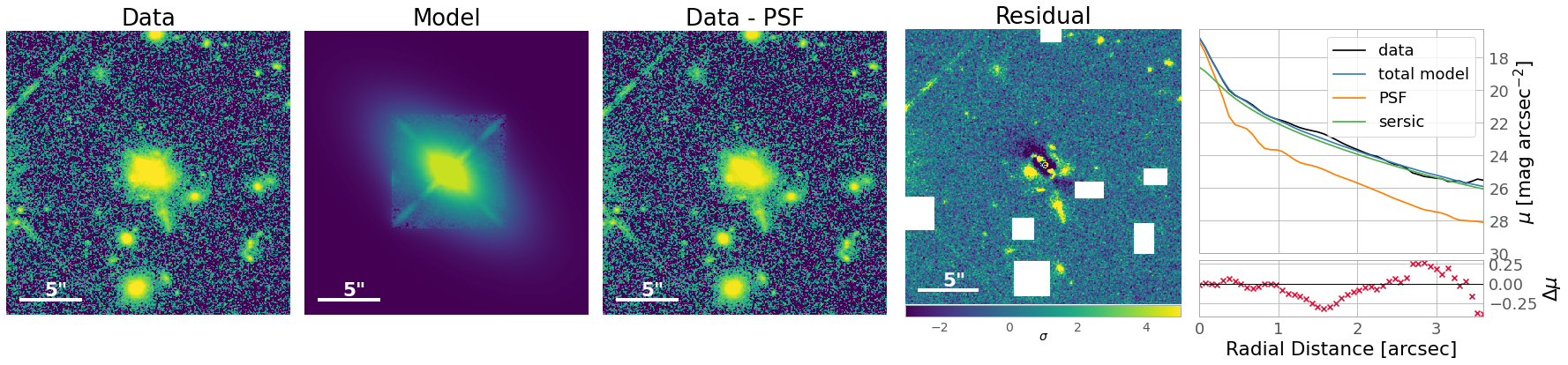}\label{fig:con14}}\\
    \subfloat[093027.21+240232.7]{\includegraphics[width=0.95\textwidth]{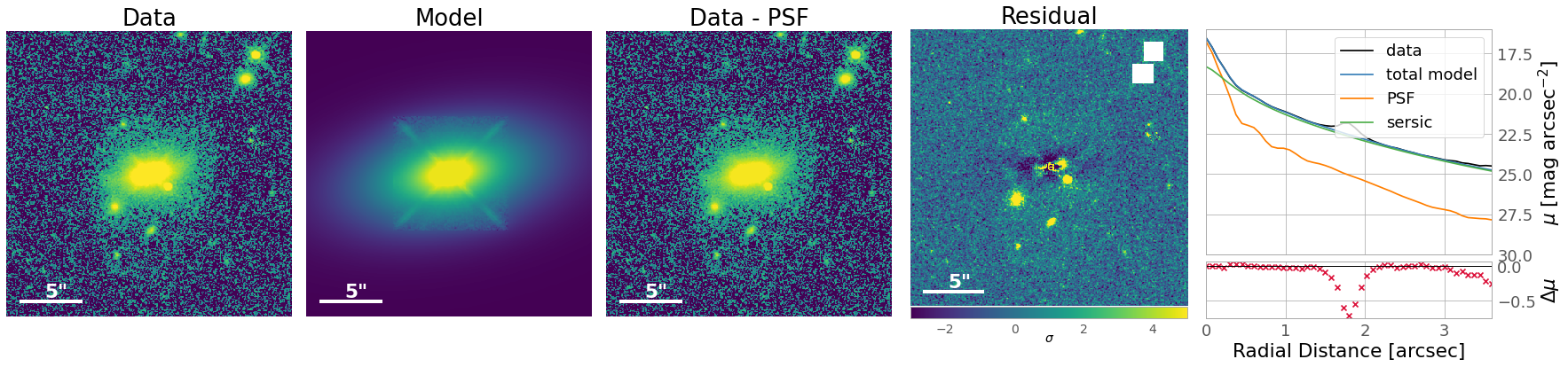}\label{fig:con25}}
    \caption{}
\end{figure*}
\begin{figure*}\ContinuedFloat
    \subfloat[111800.12+233651.5]{\includegraphics[width=0.95\textwidth]{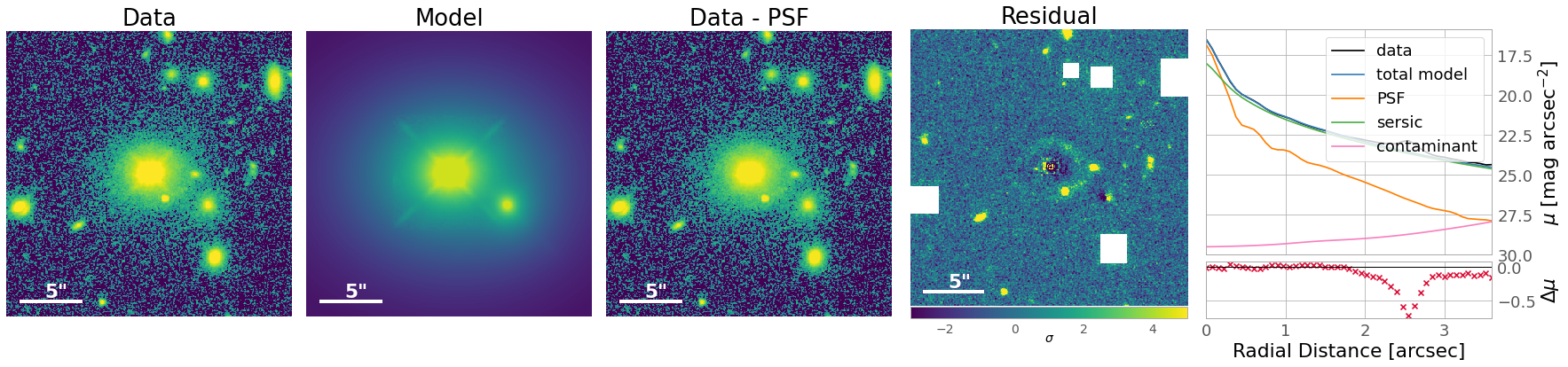}\label{fig:con40}}\\
    \subfloat[114744.11-014601.5]{\includegraphics[width=0.95\textwidth]{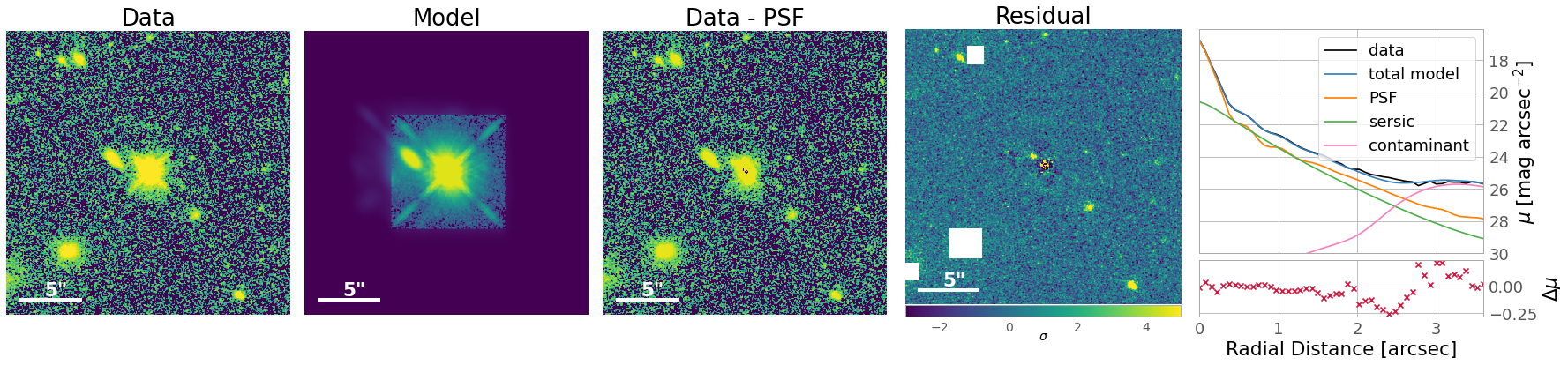}\label{fig:con44}}\\
    \subfloat[121118.66+143810.4]{\includegraphics[width=0.95\textwidth]{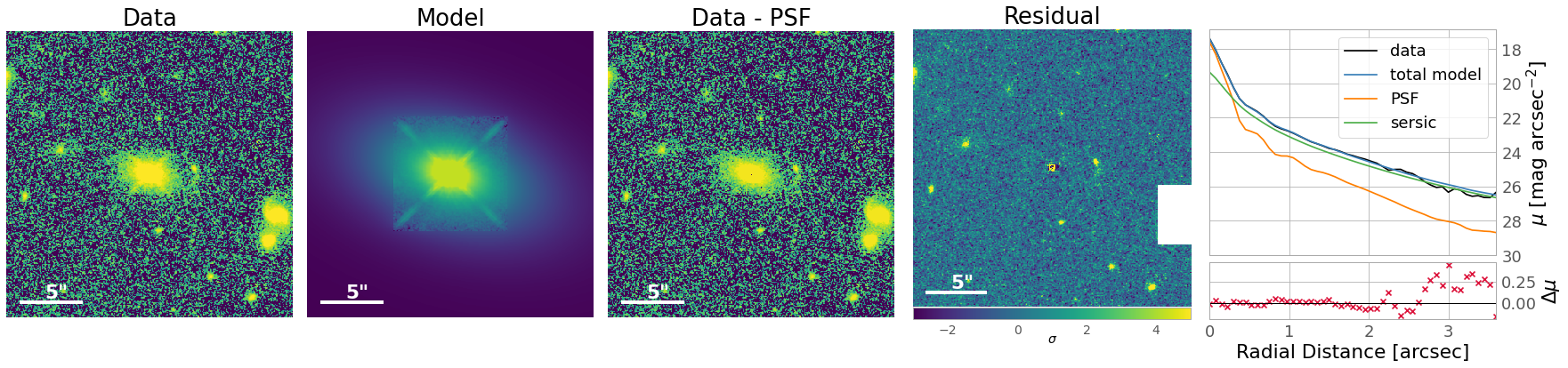}\label{fig:con48}}\\
    \subfloat[210927.29-061015.1]{\includegraphics[width=0.95\textwidth]{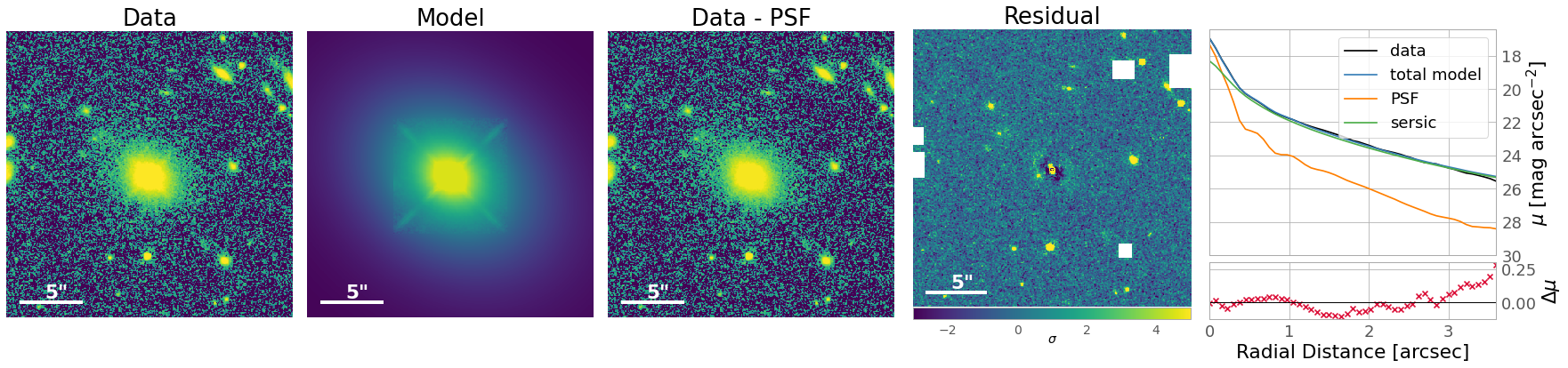}\label{fig:con85}}
    \caption{As Figure \ref{fig:soi}, but for the control sample.}
    \label{fig:con}
\end{figure*}

\subsection{Uncertainties and Limitations} \label{sec:res:ul}

It should be clearly stated that our interpretations stated here are to underscore the interesting questions posed by this sample (and control) and the need for further study and observations -- not to claim statistical confidence and representation of the larger sample of AGN host galaxies. 
The primary limitation is our small sample size, which reduces our ability to generalize to the broader population of AGN hosts, and to recognize subtle differences in host parameters between our SoI and control. Another possible contributor to our results is the reduction in accuracy described above in using a single PSF model for our SoI and control, as derived from our science observations. We intentionally have a trade-off in the types of features our \HST\ observations are sensitive to - the F160W filter observes old stellar populations, and we selected it to maximize the our sensitivity to the underlying features of the host galaxy instead of the quasar light. However, if asymmetrical features arising from the merger are formed of young stars, they may be better detected in optical than near-IR. Finally, merger features in the host may be less effective as a tracer of binary SMBH hosts if merger feature dissipation in the hosts acts on a shorter timescale than the BH mergers themselves. These problems can only be overcome with both a larger sample of interest, and a multiple-times-larger control sample which, as we have indicated above, is a time-consuming and non-trivial effort.

Our fitting results from \galfit\ are subject to additional degeneracy between the AGN PSF and the host \Sersic\ profile. Generally, \galfit\ is unable to distinguish between the light from an extremely centrally concentrated \Sersic\ profile, and that of a PSF. We overcome this degeneracy by imposing strict constraints on the relative brightness of the PSF and host profiles. \citet{peng_detailed_2010} advise against using constraints as this can cause \galfit\ to fall into local minima, but their use is unavoidable in this case.  The result is we do not claim confidence in the relative brightness of the AGN and host arising from our results, and acknowledge the extent of the \Sersic\ profile may be affected. However, these components are symmetrical, and uncertainty in their precise values should not affect differences in asymmetrical features across the samples. We rely on our PSF subtraction to mitigate a final problem - some of our merger statistics (in particular, $M_{20}$) have been shown to be biased by AGN light contributing above a threshold of $\sim20\%$ \citep{getachew-woreta_effect_2022}. This should be mitigated by our careful subtraction of AGN light, but is an additional source of uncertainty.

\section{Discussion} \label{sec:dis}

If most galaxies are thought to host SMBHs and have undergone major mergers, the lack of confirmed sub-pc binary SMBHs remains a puzzle. Our work approaches this problem by asking whether candidate binary SMBH hosts show signs of recent mergers, and we find no statistically significant evidence of such. Despite the statistical concerns described above, we see two possible causes of our null result which merit discussion and future investigation. The first is that the merger feature dissipation time scale may be much shorter than the merger time scale of SMBHs. That would mean the merger indicators of the host may have completely faded while the central BSBH is still present (i.e., not yet coalesced). On one hand, this would be consistent with the ``Final Parsec Problem'' outlined above, but on the other, this increases the difficulty of reconciling a long merger time with the lack of observational evidence of close binary SMBHs. A specific test of this tension would be a study of galaxies selected to be end-stage in a recent merger - fully merged but still with asymmetries or other tidal features which have not dissipated. If indeed the time scale of BSBH merger is much longer than that of host merger feature dissipation, then one would expect all or nearly all hosts with visible merger features to contain two SMBHs, regardless of merger stage. Previous work on structural features of post-merger galaxies \citep[][e.g.]{pawlik_shape_2016, ellison_galaxy_2013} do not investigate the presence of BSBHs in these systems, and present reasonable samples for this further investigation.

The second possibility meriting discussion is whether our sample of interest is indeed composed of binary SMBH hosts. The radial velocity selection technique relies on the consistency of RV shifts across multiple epochs and broad emission lines (e.g., \hbeta\ and \MgII), which reduces the likelihood of false positives due to stochastic BLR variability \citep{liu_constraining_2014}. However, our null result suggests that additional tests, such as high-resolution imaging or long-term monitoring, are needed to confirm the binary nature of these candidates. A larger sample of BSBH candidates, combined with high-resolution imaging and long-term spectroscopic monitoring, would provide critical insights into the co-evolution of SMBHs and their host galaxies. Such a study could resolve key outstanding questions, such as the timescales of SMBH mergers relative to galaxy mergers, the prevalence of sub-pc BSBHs, and the mechanisms driving their orbital decay. In addition, follow-up with polarimetry \citep[e.g.][using ALFOSC and VLT]{marin_polarimetry_2023} and in X-ray \citep[e.g.][using \textit{Chandra}]{mondal_detection_2024} would be invaluable and essential for truly well-constrained binary and non-binary samples.

As stated above, there is degeneracy between the central AGN and host \Sersic\ profile, which makes AGN subtraction more challenging, which consequently reduces our confidence in our host morphological parameters -- these cascading challenges are well-reflected in the literature. \citet{getachew-woreta_effect_2022} find that Gini coefficient and asymmetry remain relatively stable at AGN contributions over 20\%, but M$_{20}$ moment becomes unstable. This underscores the importance of careful AGN light subtraction as we outline. Simulation work has largely been constrained to the hosts of wider-separation BSBHs due to limits in relative scale, however \citet{saeedzadeh_dual_2024} find that binary AGN host halos generally have consistent properties with non-binary AGN hosts, likely in part due to both samples showing a large diversity. \citet{bardati_signatures_2024} argue, in accord with our findings, that none of our listed merger-related metrics can individually discriminate between post-merger and ordinary galaxies, though there is potential for a multidimensional approach -- \citet{izquierdo-villalba_properties_2023}, cited within, argues that merger features in the hosts ought to dissipate on timescales much shorter than that of SMBH inspiral and merger. Other work tends to focus on AGN activity in ongoing mergers \citep[e.g.][]{sharma_connection_2024}, and thus before central SMBHs would be considered to be in a binary.

\section{Conclusion} \label{sec:con}

Binary supermassive black holes are important in studies of gravitational waves, galaxy evolution, and cosmology, but remain undetected by direct means. In particular, BSBHs at sub-parsec separations are of interest, having overcome a theoretical stalling point at $\leq$ pc scales. Our work is centered around a sample of 8 BSBH candidates which show significant shifts in the broad \hbeta\ lines and corroborating \halpha\ or \MgII over a few years, with no significant change in the emission-line profile, indicating a BSBH with a $\sim0.1$ pc separation. These AGN were selected via repeated SDSS observations and additional follow-up spectroscopy, and we herein analyze new \HST\ F160W imaging. This sample of interest was compared to a control sample of ordinary AGN with archival F160W imaging.

After \galfit\ modelling of our SoI and control, we note a moderately elevated brightness in both PSF and \Sersic\ profile in our control sample, in contrast to their relative SDSS I-band magnitudes. In review of the merger statistics of our samples, we find that after PSF subtraction, shape asymmetry indicates 1 out of 8 galaxies in the SoI and 3 out of 8 galaxies in the control are recent mergers, and Gini-M$_{20}$ indicates 3 out of 8 in both samples are recent mergers. These statistics do not indicate an elevated rate of mergers in our candidate BSBH hosts, and the K-S test and Mann-Whitney U test do not indicate our samples come from distinct populations (with one borderline result in PSF-subtracted shape asymmetry). After visual inspection, we note extended asymmetrical structure in 5 out of 8 of our SoI hosts, with none at the same degree of clarity in our control sample (though there are fainter potential features in 4 out of 8). These results and further interpretation are restricted by our small sample size, use of a single PSF for both samples due to practical restrictions, and degeneracy between the central PSF and host \Sersic\ profile in model fitting.

An open question posed by our results is whether we should expect to see faint merger features in hosts of sub-pc BSBHs -- it is possible that the dissipation timescale of merger features in the host may be significantly shorter than the merger timescale of the SMBHs. It is also reasonable in the light of our null result to give additional consideration to the BSBH nature of our sample of interest, but radial velocity shifts across multiple epochs (as seen in our sample) are unlikely to be explained in a non-binary SMBH system. A study of post-merger galaxies with still-extant merger features would be highly effective at resolving some of these open questions, as our result would indicate such galaxies should still retain the two separate SMBHs from their progenitors. In addition, clear improvements could be made by a larger sample size in both targets and control, though this can be technically challenging.

\section{Data Availability}

All \HST\ data used in this paper can be found via MAST: \dataset[10.17909/abx2-fd14]{http://dx.doi.org/10.17909/abx2-fd14}.

\section*{Acknowledgments}

L.N. thanks A. Gross and M. Verrico for advice throughout this study.
The authors thank A. Vick for their help with our HST observations. Support for Program number HST-GO-15975 (PI: X. Liu) was provided by NASA through grants from the Space Telescope Science Institute, which is operated by the Association of Universities for Research in Astronomy, Incorporated, under NASA contract NAS5-26555. L.N., X.L., and Y.C.C., acknowledge support from the University of Illinois Campus Research Board and NSF grant AST-2206499. This material is based upon work supported by the National Science Foundation Graduate Research Fellowship Program under Grant No. AJ840. Any opinions, findings, and conclusions or recommendations expressed in this material are those of the author(s) and do not necessarily reflect the views of the National Science Foundation. Y.S. acknowledges support from NSF grant AST-2009947. An early version of this work was presented as an iPoster at the 243rd meeting of the American Astronomical Society.


\bibliography{fabulovs22}{}
\bibliographystyle{aasjournalv7}



\appendix

\section{\galfit\ Example Files} \label{app:gal}

\textbf{\galfit\ Input}

\begin{verbatim}
===============================================================================
# IMAGE and GALFIT CONTROL PARAMETERS
A) image.fits      # Input data image (FITS file)
B) galfit.fits     # Output data image block
C) None            # Sigma image name (made from data if blank or "none")
D) psf.fits        # Input PSF image and (optional) diffusion kernel
E) 3               # PSF fine sampling factor relative to data
F) mask.fits       # Bad pixel mask (FITS image or ASCII coord list)
G) constraints     # File with parameter constraints (ASCII file) 
H) 1 400 1 400     # Image region to fit (xmin xmax ymin ymax)
I) 400 400         # Size of the convolution box (x y)
J) 28.178874       # Magnitude photometric zeropoint 
K) 0.06 0.06       # Plate scale (dx dy)   [arcsec per pixel]
O) regular         # Display type (regular, curses, both)
P) 0               # Choose: 0=optimize, 1=model, 2=imgblock, 3=subcomps
    
 0) sky
 1) 305.90    0        # sky background       [ADU counts]
 2) 0.00      0        # dsky/dx (sky gradient in x) 
 3) 0.00      0        # dsky/dy (sky gradient in x) 
 Z) 0                  #  Skip this model in output image?  (yes=1, no=0)
    
 0) psf                # object type
 1) 199.66 200.91  1 1  # position x, y        [pixel]
 3) 20.27     1        # total magnitude
 Z) 0                  #  Skip this model in output image?  (yes=1, no=0)
 
 0) sersic             # object type
 1) 199.66 200.91  1 1  # position x, y        [pixel]
 3) 20.27     1        # total magnitude
 4) 3.90     1        # length
 5) 2.00      1        # index
 9) 0.84      1        # axis ratio
 10) 0.00    1        # PA
 Z) 0                  #  Skip this model in output image?  (yes=1, no=0)
\end{verbatim}

\textbf{Constraint File}

\begin{verbatim}
# Component/    parameter   constraint  Comment
# operation    (see below)    range

# Component 2: psf
      2           x          -2 2      #
      2           y          -2 2      #
      2           3          -5 5      # mag

# Component 3: sersic
      3           x          -2 2      #
      3           y          -2 2      #
      3           3          -5 5      # mag
      3           4          5 to 50   # length
      3           5        0.3 to 8    # index
      3           9        0.1 to 1    # axis ratio}
\end{verbatim}



\label{lastpage}
\end{document}